\documentclass[paper,11pt]{JHEP}
\usepackage[centertags]{amsmath}
\usepackage{amsfonts} \usepackage{amssymb} \usepackage{amsthm}
\allowdisplaybreaks[1]
\usepackage{graphicx}
\usepackage{caption}

\newcommand{\bra}[1]{\langle{#1}|}
\newcommand{\ket}[1]{|{#1}\rangle}
\newcommand{\vev}[1]{\langle{#1}\rangle}
\def\one{{\rm 1\kern -.9mm l}}                             %
\newcommand{\braket}[2]{\langle{#1}|{#2}\rangle}
\def\beq{\begin{equation}}
\def\eeq{\end{equation}}
\def\beq{\begin{equation}}
\def\eeq{\end{equation}}
\def\beqa{\begin{eqnarray}}
\def\eeqa{\end{eqnarray}}
\newcommand{\eqa}{\begin{eqnarray}}
\newcommand{\ena}{\end{eqnarray}}
\newcommand{\eq}[1]{eq. (\ref{#1})}

\def\gh{{\rm gh}}
\def\ii{\mathrm{i}}
\def\ee{\mathrm{e}}
\newcommand{\Z}{\mathbb{Z}}

\newcommand{\comm}[2]{\left[ #1,#2\right]}

\newcommand{\cale}{\mathcal{E}}
\newcommand{\cA}{\mathcal{A}}
\newcommand{\cE}{\mathcal{E}}
\newcommand{\cL}{\mathcal{L}}

\newcommand{\cW}{\mathcal{W}}
\newcommand{\Bop}{B_{\mathrm{op}}}
\newcommand{\Rw}{R_{\mathrm{w}}}

\title{New numerical results and novel effective string predictions  for
Wilson loops%
}
\author{M. Bill\'o, M. Caselle, R. Pellegrini
\\
\vskip 0.2cm
Dipartimento di Fisica Teorica, Universit\`a di Torino\\
and Istituto Nazionale di Fisica Nucleare - sezione di Torino \\
Via P. Giuria 1, I-10125 Torino, Italy\\
\vspace{0.25cm}
\email{billo,caselle,pellegri@to.infn.it} 

}
\abstract{We compute the prediction of the Nambu-Goto effective string model
for a rectangular Wilson loop up to three loops. This is done through the use of
an operatorial, first order formulation and of the open string analogues of 
boundary states. This result is interesting since there are universality theorems stating
that the predictions up to three loops are common to all effective string models.
To test the effective string prediction,
we use the Montecarlo evaluation, in the 3d Ising gauge model, of an 
observable (the ratio of two Wilson loops with the same perimeter) for which
boundary effects are relatively small. Our simulation attains 
a level of precision which is sufficient to test the two-loop correction.
The three-loop correction seems to go in the right direction, but is actually yet 
beyond the reach of our simulation, since its effect is comparable with
the statistical errors of the latter.
}
\preprint{DFTT/18/2011}

\begin{document}

\section{Introduction}
\label{sec:intro}
The proposal that the strong-coupling dynamics of gauge theories could be
captured by an effective string theory is a long-standing one
\cite{Wilson:1974sk}-\nocite{Mandelstam:1974pi,Nielsen:1973cs,'tHooft:1974hx,
Polyakov:1979gp,Polyakov:1980ca,Nambu:1974zg,Nambu:1978bd,Luscher:1980fr}\cite{
Luscher:1980ac}; the fluctuating long string, in this approach, is the color
flux tube joining coloured charges. Perhaps the most direct prediction that can
be extracted from the effective string is the shape of the potential $V(R)$
between two external sources (two static quarks) at distance $R$. From
the gauge theory point of view, if we have a rectangular Wilson loop $W(L,R)$ of
sides $L$ and $R$, the inter-quark potential is
\begin{equation}
\label{potdef}
V(R)=-\lim_{L\to\infty} \frac{1}{L}\log W(L,R)~.
\end{equation}
In the confining phase, the area law for the Wilson loop corresponds to a
linear potential 
\begin{equation}
\label{Varea}
V(R) = \sigma R + \ldots~.
\end{equation}
 In a string interpretation, the
area term $\sigma L R$ in the exponent of the Wilson loop is the classical
action of an effective string model. Another way of measuring the static $q\bar q$
potential is to consider the correlator of two Polyakov loops at distance $R$ 
in a gauge theory at finite temperature $\propto 1/L$; in the strong-coupling regime,
the string world-sheet describing this situation is a cylinder. 

The simplest and most obvious choice for the effective string is the Nambu-Goto
one \cite{nambu1970,Goto:1971ce}, where the action is the induced area of the
string world-sheet, see \eq{ngaction} below, and  
%The invariance of this action under
%re-parametrizations of the world-sheet can be exploited to reach a ``physical
%gauge'' in which the world-sheet is spanned by two of the $d$ space-time
%coordinates (e.g., the ones pertaining to the plane where the Wilson loop lays);
the degrees of freedom are the $D=d-2$ transverse fields.
% $\vec X(x^1,x^2)$.
The action is non linear and contains, besides the kinetic term, a series of
higher derivative interactions. Quantization is implemented by functional
integration and can be carried out perturbatively. This approach was used in
$\cite{Dietz:1982uc}$ to compute, up to two loop order, the string vacuum
amplitude on world-sheets with disk, cylinder or torus topology (the first two
being relevant for Wilson loops and Polyakov loop correlators respectively). 

On general grounds, one expects any effective theory for a long string
to contain as degrees of freedom the transverse fields, representing the
Goldstone modes for the translational invariance 
broken by the string configuration. The ``one loop'' quantum corrections are
then given by the functional determinants of the kinetic operators for the
transverse modes and are the same for all effective models. They lead to the
correction to the static potential known as ``L\"uscher
term''~\cite{Luscher:1980fr,Luscher:1980ac}:
\begin{equation}
 \label{luescher}
V(R)= \sigma R - \frac{D\pi}{24 R} + \ldots~.
\end{equation}
Over the years, computer simulations of Polyakov loop correlators and Wilson
loops have shown with increasing evidence the presence of this correction (for
an up to date review see \cite{Teper:2009uf}), confirming the soundness of the
effective string approach.

The various possible effective theories are distinguished by the form of their
interaction terms, which in turn affect the perturbative corrections to the
amplitudes, starting at two loops. The interaction terms cannot, however, be
completely arbitrary. In \cite{Luscher:2004ib} it was shown that the requirement
that the effective string cylinder partition function (corresponding to a
Polyakov loop correlator) can be re-expressed in terms of propagating closed
string states fixes the coefficients of the first higher derivative terms.
Moreover, if the effective string is supposed to describe the low-energy regime
of a consistent quantum field theory, the Poincar\'e symmetry of the latter
must be still realized. Since the configuration around which the long string
fluctuates spontaneously breaks, in particular, the Lorentz transformations
mixing longitudinal and transverse coordinates, these must be realized
non-linearly in the effective action. This requirement is present in the
analysis of \cite{Luscher:2004ib}, see \cite{Meyer:2006qx}, but was made more
explicit and pursued further in
\cite{Aharony:2009gg}-\nocite{Aharony:2010cx}\cite{Aharony:2010db}, where it is
shown that the coefficients of the quartic and sextic derivative interaction
terms are almost completely fixed by this constraint, both for closed and open
strings, and coincide with those of the NG model. This ''universality'' of the
lowest interaction terms suggests that Montecarlo simulations
should be compatible with NG prediction also beyond the one-loop level. 

Testing the effective string predictions beyond the one-loop terms, for instance
by looking at subleading terms in the $1/R$ expansion of the effective potential,
is a very serious challenge. It is a challenge from the computational point of
view, because it requires to carry out simulations so precise as to distinguish
deviations of the results from the one-loop predictions. However, thanks to the
development of new algorithms and the surge in available computer power, such a
precision is basically within reach. In particular, in the case of the three
dimensional gauge Ising model in which, exploiting duality, high precision can be
reached at a reasonable computational cost, universality was unambiguously
confirmed at two loop level both in the case of the torus
geometry~\cite{Caselle:1994df,Caselle:2007yc,Caselle:2006dv} and in the case of
the cylinder geometry~\cite{Caselle:2002ah,Caselle:2005xy}. Testing universality
at three loop level turned out to be more difficult. The only
existing tests~\cite{Caselle:2010pf} were performed in very asymmetric
geometries were string effects are magnified but systematic errors, due for
instance to the vicinity of the deconfinement transition, are more difficult to
control. For a survey of most recent results for other gauge groups or other
observables see~\cite{Teper:2009uf}). Until now no attempt to go beyond the one
loop results was performed in the case of the disk geometry. Filling this gap is
one of the main goals of the present paper. 

Also on the theoretical side, extracting the higher loop effects of the
effective string model is not an easy task. The diagrammatic evaluation of the
effective string amplitudes
% employing $\zeta$-function regularization
carried out in \cite{Dietz:1982uc}
% up to two loops 
was pushed up to three loops, and extended to a generic effective string action
to this order in \cite{Aharony:2009gg}, for the cylinder partition function only.

In this work, we derive higher loop corrections
in the Nambu-Goto disk amplitude corresponding to the Wilson loop. Our method,
which allows in fact to obtain the exact amplitude that resums the loop
expansion, is based on the 
%operatorial evaluation of 
the first-order
reformulation of the Nambu-Goto string model; it has already been
successfully applied to the cylinder \cite{Billo:2005iv} and the torus
\cite{Billo:2006zg,Billo:2007fm} partition functions. We will see that the
Wilson loop case is more delicate, but we propose an exact expression which
reproduces, at two-loop, the result of \cite{Dietz:1982uc}: this represents a
very non-trivial check and we trust therefore our results also to higher loops. 

The universality of the first terms in the derivative expansion of the
effective action explains why the NG model represents a good
approximation of the correct effective string theory for the flux tube; in turn,
testing NG to the order to which its effects are universal would represent a 
stringent test of the whole effective string approach.

Still, the NG model can not be the right one, and deviations from its
predictions, when precisely identified, may help in the quest for the correct 
effective string. In particular, the Lorentz
invariance should be preserved also at the quantum level. The quantum
consistency of the NG theory is usually investigated  by using a first-order
re-formulation, which involves an independent metric on the world-sheet but is
quadratic in the string fields: employing the Weyl invariance to reach the
so-called ''conformal gauge'', the world-sheet metric decouples and one is left
with a quadratic action, to be supplemented with the Virasoro constraints. This
is the standard approach described in string theory textbooks, see for instance
\cite{Green:1987sp,Polchinski:1998rq}. The residual conformal invariance
generated by the modes of the Virasoro constraints can be exploited to fix a
light-cone (physical) gauge where only the $D$ transverse fields are
independent; it can also be enforced, in the ``covariant quantization'', 
by taking into account the effects of the
ghost/anti-ghost system that arise from the Jacobian to fix the conformal gauge.
In the light-cone gauge, the closure of the target-space Lorentz algebra at the
quantum level requires $D=24$; however, when the string fluctuates around
a long string configuration with length scale $R$, this anomaly is
suppressed in the large $R$ limit \cite{Olesen:1985pv}. In covariant
quantization, the central charge $c_{\mathrm{gh}}=-26$ of the ghost/anti-ghost
system has to be compensated by that of $D+2=26$ bosonic fields. 

Polyakov's approach \cite{Polyakov:1981rd} of functionally integrating over the
intrinsic metric shows that for $D\not=26$ the scale of this metric does not
decouple and represents an additional degree of freedom that has to be taken
into account to ensure quantum consistency. This world-sheet boson is usually
referred to as the Liouville field since its action is of the Liouville type. 
If the effective string theory must contain just the $D$
transverse d.o.f., and realize also at the quantum level the
Lorentz algebra, none of the standard quantizations can thus be correct.
An interesting proposal was put forward by Polchinski and Strominger in
\cite{Polchinski:1991ax}; to reconcile the requirement of critical central
charge with having just the $D$ transverse degrees of freedom, they proposed
a Polyakov model where the Liouville field is re-expressed in terms of the
induced metric, i.e., in terms of the transverse fields. The action becomes
non-local, but admits nevertheless a sensible expansion around long string
configurations. It has been shown that the first terms in the derivative
expansion of this model agree with the NG ones, in accordance with the
universality arguments. It seems however difficult to compute higher loop
corrections to the amplitudes in this model. 

Recently, some particular, supersymmetric, gauge theories have been finally
given an explicit realization in terms of strings propagating on a curved
manifold with an extra dimension, in the so-called AdS/CFT duality
\cite{Maldacena:1997re,Gubser:1998bc,Witten:1998qj}; the extra dimension can be
interpreted in terms of the energy scale \cite{Polyakov:1998ju} and as such it
does not represent a spurious degree of freedom. From the point of view of the
effective action for the transverse fields, the interaction terms would take
into account the curved geometry. This is of course a very intriguing
possibility.

There is an extra issue in devising the correct effective string action, namely the 
possible presence of boundary terms. Lattice simulations have made it clear 
that the Wilson loop expectation value displays, beside the leading area term, a 
perimeter term with an independent coupling:
\begin{equation}
\label{perimeter}
W(L,R) \sim \exp\left(-\sigma LR - \mu (L+R) + \ldots\right)~.
\end{equation}
In this paper, we will study a particular observable, corresponding to the ratio
of two different rectangular Wilson loops, such that the leading perimeter
dependence described in \eq{perimeter} cancels \cite{Caselle:1996ii}. However, it is natural
to expect that the perimeter term is only the classical value of some boundary
component of the effective action which also yields other, subleading,
corrections to the amplitudes.
In \cite{Luscher:2004ib,Aharony:2010cx,Aharony:2010db} the first few possible
boundary terms in a derivative expansion of the effective action have been
investigated for the Polyakov loop geometry, finding that they can lead to
corrections of the order of $1/R^4$ ($R$ being the distance between the
Polyakov loops); to our knowledge, no explicit evaluation of the effect of boundary terms in the
Wilson loop geometry is available. From the theoretical point of view, 
in  \cite{Durhuus:1981gu}-\nocite{Durhuus:1981ad,Durhuus:1982fd,DiVecchia:1982bz}\cite{Alvarez:1982zi}
by considering Polyakov quantization in presence of boundaries, it was found that
boundary terms in the world-sheet actions are needed, and in particular a term
proportional to the induced length of the boundary, whose leading contribution to 
the Wilson loop amplitude would indeed be a perimeter term. It would be nice to 
investigate the modifications of the Nambu-Goto calculations in presence of such a 
term, but we leave this to future work.  

In comparing the outcomes of our simulations to the theoretical predictions
of the Nambu-Goto effective model, we will have to take into account the 
likely presence of boundary terms, of which no explicit evaluation is available; 
this increases the uncertainties in the comparison.  

As mentioned above the main goal of this paper was to compare theoretical
predictions and Montecarlo simulations beyond one loop in the case of the Wilson
loop geometry. We shall show that, using duality based algorithms, it is
possible to study Wilson loops large enough to make boundary effects negligible
and at the same time to reach a precision high enough to disentangle effective
string corrections beyond the one loop level. As in the cases of the torus and
cylinder geometry, we shall be able to confirm universality at the two loops
level. The third loop corrections seem to go in the right
direction, but their magnitude is of the same order as the 
statistical errors, so that they remain beyond the reach of present simulations.

\section{Two-loop effective string prediction}
\label{sec:twoloop}
In this section we summarize the two-loop prediction for the Wilson loop amplitude
obtained in the Nambu-Goto effective string model. As discussed 
in the introduction, the predictions up to three loops are actually universal 
for all effective string models. 

In the Nambu-Goto approach \cite{nambu1970,Goto:1971ce} the action for the fluctuating
surface spanned by the flux tube is simply its
induced area in the $d$-dimensional target space:
\begin{equation}
\label{ngaction}
S = \sigma \int d^2\xi \sqrt{\det g}~,
\hskip 0.6cm g_{\alpha\beta} = \frac{\partial
X^M}{\partial\xi^\alpha}\frac{\partial X^N}{\partial \xi^\beta} G_{MN}~.
\end{equation}
Here $\sigma$ is the string tension, the surface is parametrized by proper
coordinates $\xi^\alpha$, and
$X^M(\xi)$ ($M=1,\ldots,D+2$) describe the target space position of a point
specified by $\xi$.
For us the target space metric $G_{MN}$ will always be the flat one. 
Our analysis aims at comparing effective string predictions with lattice
simulations of gauge theories, which are always performed with an
Euclidean time. As a consequence, we use the Euclidean signature on the 
target space as well as on the world-sheet; this is what is commonly done in the  literature about the
effective string approach 

Consider a rectangular Wilson loop of sides $L$ and $R$ in the $x^1, x^2$ plane. 
The direction $x^1$ is interpreted as the Euclidean time, so that in the limit 
$L\to \infty$ one can extract the static quark-antiquark potential via \eq{potdef} 
and compare it directly with lattice simulations. 

In the so-called ``static'' gauge the proper coordinates of the string are
identified with $x^1$ and $x^2$ already at the classical level.
The partition function for the  Wilson loop surface is then obtained by
functional integration over the transverse fields $\vec X(x^1,x^2)$: 
\begin{equation}
\label{ngpart}
\begin{aligned}
W(L,R) & = \int D\vec X\, \exp\left\{-\sigma
\int_0^L dx^1 \int_0^R dx^2 \left(1 + (\partial_1\vec X)^2
 + (\partial_2\vec X)^2 + (\partial_1 \vec X \wedge \partial_2 \vec
X)^2\right)^{\frac 12}\right\}
\\
& = \int D\vec X\,  \exp\left\{-\sigma 
\int_0^L dx^1 \int_0^R dx^2 \left[
1 +\frac 12 (\partial_1\vec X)^2 + \frac 12 (\partial_2\vec X)^2 +
\mbox{interactions} \right]\right\}~,
\end{aligned}
\end{equation}
with the transverse fields $\vec X$ vanishing on the Wilson loop
perimeter. To this NG action, terms living on the boundary of the domain can
(and must) be added, and their effect is expected to be important, especially
for small loops. In this paper, however, we will not deal explicitly with the
quantum corrections induced by such terms. Eq. \ref{ngpart} describes a statistical 
sum over surfaces bounded by the Wilson loop.

By construction, we must have
\begin{equation}
\label{invW}
 W(L,R) = W(R,L)~.
\end{equation}

Expanding the square root as in the second line of \eq{ngpart}, the constant term
leads to the well-known classical area law
\begin{equation}
W_{\mathrm{cl}}(R,L)=\ee^{-\sigma RL+p(R+L)+k}~.
\label{wl1}
\end{equation}
Here on top of the area law we have included a perimeter term, whose
strength is parametrized by $p$, which arises from boundary terms in the action.
The overall normalization, which is out of control, is parametrized by $k$.

{}From the functional integration over the fields $\vec X$, with the bulk action
given in the second line of \eq{ngpart}, one gets the determinant of their
kinetic operator multiplied by perturbative corrections, starting at two loops.
The result has the structure
\begin{equation}
 \label{wlngexp}
W(L,R) = \ee^{-\sigma \cA+p(R+L)+k}\, \left(\frac{\eta(\ii
u)}{\sqrt{R}}\right)^{-\frac D2}\left\{1 + \frac{\cL_2(u)}{\sigma\cA} +
\frac{\cL_3(u)}{(\sigma\cA)^2} + \ldots\right\}~,
\end{equation}
where, for ease of notation, we have introduced
\begin{equation}
\label{defAu}
\cA = R L~,~~~ u = \frac{L}{R}~.
\end{equation}
In \eq{wlngexp}, the term $(\eta(\ii u)/\sqrt{R})^{D/2}$, where by $\eta$ we
denote Dedekind's function (see Appendix \ref{app:not} for relevant
notation and definitions) is the functional determinant; it is invariant under
$R\leftrightarrow L$, i.e., under $u \leftrightarrow 1/u$, as it follows from
the modular transformation properties of the $\eta$ function. The corresponding
contribution to the interquark potential of \eq{potdef}
is the celebrated L\"uscher term
$D\pi/(24 R)$ appearing in \eq{luescher} that 
represents the leading correction to  the linear confining
potential~\cite{Luscher:1980fr,Luscher:1980ac}.

With the impressive enhancement in precision of Montecarlo simulations in the
last years and, above all, in view of the universality theorems
discussed in the introduction it becomes important to evaluate and test also the
subleading terms in \eq{wlngexp}. The terms $\cL_a(\ii u)$ represent the
corrections at order $a$ in the loop expansion parameter $\sigma\cA$;
they must be invariant under $u\leftrightarrow 1/u$.

The computation of the two-loop%
\footnote{In fact, \cite{Dietz:1982uc} considers the three cases where the
string world-sheet is a a disk, a cylinder or a torus, relevant for the
effective description of, respectively, Wilson loops, Polyakov loop correlators
and interfaces in a compact target space.} correction $\cL_2$ was carried out
almost 30 years ago by Dietz and Filk \cite{Dietz:1982uc}. 
However, the two-loop result of \cite{Dietz:1982uc} appears to be incompatible
with the data obtained with various simulation techniques, see
\cite{Billo':2010ix} for an account of recent attempts in this direction. This
fact prompted us to a critical re-examination of the Dietz and Filk calculation.
We found a mistake%
\footnote{The term proportional to $\lambda_2$ in that equation should read 
$D(\vev 3+\vev 4+2\vev 1+4 \vev 2)$ instead of $D(\vev 3+\vev 4+2\vev 1-4\vev 2)$. 
This mistake was not easy to spot because its contribution vanishes in the
cylinder and torus geometries; in these cases, in fact, the two loop prediction of
Dietz and Filk is compatible with the Montecarlo simulations
\cite{Caselle:1994df,Caselle:2007yc,Caselle:2002ah}.
Moreover the term affected by this error is symmetric under the exchange 
$R\leftrightarrow L$ and thus can not be excluded
using this symmetry requirement.} in their eq. (3.3).
The correct result is
\begin{equation}
\cL_2(u)
=\left(\frac{\pi}{24}\right)^2 \left[2D\, u^2 E_4\left(\ii u\right)- 
\frac{D(D-4)}{2} E_2\left(\ii u\right)E_2\left(\ii/u\right) \right]~.
\label{2lng}
\end{equation}
where $E_2$ and $E_4$ denote the second and fourth order Eisenstein
series, see Appendix \ref{app:not}.
We shall show in the next section that \eq{2lng} turns out to be
compatible with the Montecarlo results%
\footnote{Note however that, as said before, the presence of boundary terms
in the string action leads, besides the classical perimeter term, to further
quantum corrections to \eq{wlngexp}, which we are not taking into account in our
theoretical predictions.}, in accordance with its universal character.

\section{New numerical results}
\label{sec:numerical}
In this section we compare the two-loops effective string prediction for Wilson loops
with a set of numerical data extracted from a Montecarlo simulation of 
a suitable observable in the three-dimensional $ \Z_2 $ gauge model.
\subsection{Definition of the observable  in the 3d  $ \Z_2 $ gauge model}
\label{subsec:defobs}
The 3d $ \Z_2 $ gauge model on a cubic lattice is defined by the partition function
\beq
Z_{g}(\beta)=\sum_{\sigma_l=\pm 1} \exp \{-\beta S_g\}~.
\eeq
The action $ S_g $ is a sum over all the plaquettes of the lattice,
\begin{equation}
S_{g}=-\sum_{\Box} \sigma_{\Box}~,~~~ 
\sigma_{\Box}=\sigma_{l_1} \sigma_{l_2} \sigma_{l_3} \sigma_{l_4}~,
\end{equation}
where $ \sigma_l $ are Ising variables located on the links of the lattice.
This gauge model is equivalent to the standard Ising model by the 
Kramers-Wannier duality transformation
\begin{align}
\label{ampli1}
Z_{g}(\beta) & \propto Z_{s}(\tilde{\beta}) ~, 
\\ \tilde{\beta} & =-\frac{1}{2}\log[\tanh(\beta)]~,
\end{align}
where $ Z_s $ is the partition function of the Ising model in the dual lattice
\beq
Z_{s}(\tilde{\beta})=\sum_{s_i=\pm 1} \exp \{-\tilde{\beta} H_1(s)\}~.
\eeq
Here
\beq
H_1(s)=-\sum_{\langle ij \rangle} J_{\langle ij \rangle} s_i s_j
\eeq
with $ i $ and $ j $ denoting the nodes of the dual lattice and the sum is over
all the links $ \langle i j \rangle $ of the dual lattice. Here the couplings $
J_{ij} $ are fixed to the value $ +1 $ for all the links. This define a
one-to-one mapping between the free energy densities in the thermodynamic limit.
The dual of the Wilson loop expectation value , in which we are interested, is
given by
\beq
W(R,L)=\frac{Z_{s,W}(\tilde{\beta})}{Z_{s}(\tilde{\beta})} ~, 
\label{new1}
\eeq
where in $ Z_{s,W} $ all the couplings of the links that intersect the surface $
S $ enclosed by the Wilson loop take the value $ J_{\langle ij \rangle}=-1 $.

\begin{figure}
\def\svgwidth{9cm}
\begin{center}
\begingroup
  \makeatletter
  \providecommand\color[2][]{%
    \errmessage{(Inkscape) Color is used for the text in Inkscape, but the package 'color.sty' is not loaded}
    \renewcommand\color[2][]{}%
  }
  \providecommand\transparent[1]{%
    \errmessage{(Inkscape) Transparency is used (non-zero) for the text in Inkscape, but the package 'transparent.sty' is not loaded}
    \renewcommand\transparent[1]{}%
  }
  \providecommand\rotatebox[2]{#2}
  \ifx\svgwidth\undefined
    \setlength{\unitlength}{264.45306702pt}
  \else
    \setlength{\unitlength}{\svgwidth}
  \fi
  \global\let\svgwidth\undefined
  \makeatother
  \begin{picture}(1,0.72971954)%
    \put(0,0){\includegraphics[width=\unitlength]{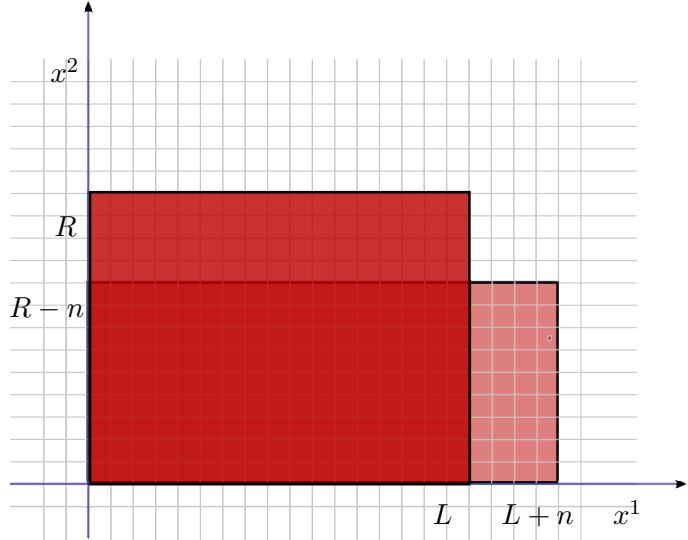}}%
    \put(0.05899263,0.6812748){\makebox(0,0)[lb]{\smash{$x^2$}}}%
    \put(-0.0015096,0.33393314){\makebox(0,0)[lb]{\smash{$R-n$}}}%
    \put(0.06504285,0.45286319){\makebox(0,0)[lb]{\smash{$R$}}}%
    \put(0.62394037,0.02790584){\makebox(0,0)[lb]{\smash{$L$}}}%
    \put(0.72561645,0.02790584){\makebox(0,0)[lb]{\smash{$L+n$}}}%
    \put(0.89025404,0.02790584){\makebox(0,0)[lb]{\smash{$x^1$}}}%
  \end{picture}%
\endgroup
\end{center}
\caption{The observable $\Rw(L,R,n)$ is given by the ratio of the Wilson loops
$W(L,R)$ and $W(L+n,R-n)$.}
 \label{fig:observable}
\end{figure}

The form of the functional dependence of the Wilson loop on the sides $R,L$ is 
given in \eq{wlngexp}, where we have to fix $D=d-2=1$.
The values of the parameters $ \sigma $,$ p $ and $ k $ appearing in the classical
part of this expression are  not predicted by the effective string theory. 
However $ p $ and $ k $ can be eliminated by considering
ratios of Wilson loops with the same perimeter \cite{Caselle:1996ii}, as the
ones of fig. \ref{fig:observable}.
We will therefore consider the observable
\beq
\label{Rw}
\Rw(L,R,n)=\frac{W(L,R)}{W(L+n,R-n)}~.
\eeq

\subsection{Algorithm and simulation settings}
\label{subsec:alg}
In general, computing large Wilson loops is a difficult task because the Wilson
loop expectation value decreases exponentially with the area, so the
relative error increases exponentially.
This is true also for the gauge Ising model but in this case, if one evaluates
the dual observable eq.(\ref{new1}), 
the problem can be overcome. Our approach is essentially the same
described in great detail in \cite{Caselle:2002ah}, adapted to 
the observable \ref{Rw}.
In order to compute \eq{Rw} we factorize the ratio as follows:
\beq
\label{factorization}
\Rw(L,R,n)=\frac{W(L,R)}{W(L+1,R)} \frac{W(L+1,R)}{W(L+2,R)} 
\dots \frac{W(L+n,R)}{W(L+n,R-1)} \dots \frac{W(L+n,R-n+1)}{W(L+n,R-n)}
\eeq
Every ratio of \eq{factorization} is then simulated separately, using a
hierarchical algorithm as explained in \cite{Caselle:2002ah}.

In order to compare the effective string theory prediction with the numerical
data we need a very precise estimate of the zero temperature string tension. We choose to simulate the model at coupling $ \beta=0.75180 $, for which the string
tension is known with very high precision \cite{Caselle:2004jq} to be $\sigma=0.0105241(15)$. We performed our simulations on a lattice of size $64\times 120\times 120$ and measured only Wilson loops lying along the directions of size $120$.
The lattice sizes were chosen large enough to avoid finite size effects; in particular, we carefully tested that a lattice size of $64$ lattice spacings in the transverse direction was enough to avoid finite size corrections even for the largest loops that we studied. 

% Table (\ref{simulation}) summarizes the basic technical information about our
% simulation.
% \begin{table}[ht]
% \centering
% \begin{tabular}{|ccc|}
% \hline
% $ \beta $ & $ \tilde{\beta}  $ &  $ \sigma $  \\
% \hline
% 0.75180 & 0.226104 & 0.0105241(15) \\
% \hline
% \end{tabular}
% \caption{Parameters used in the simulation.}
% \label{simulation}
% \end{table} 

\subsection{Comparing the results to the two-loop predictions}
\label{subsec:comp2l} 
We extracted from the simulation the ratio of expectation values
\beq
\label{Rwagain}
\Rw(L,\frac{L}{u},n)=\frac{W(L,\frac{L}{u})}{W(L+n,\frac{L}{u}-n)} 
\eeq
for different values of the asymmetry parameter $ u=L/R > 1 $ and of $ n $.
The numerical results of the simulation are reported in Table (\ref{results}).
The leading classical behaviour of the observable 
$ \Rw(L,L/u,n) $ is given by
\beq
\label{Rwcl}
\Rw(L,\frac{L}{u},n)=\exp \left\{-\sigma n\left(n+L(1-\frac{1}{u})\right) \right\}.
\eeq
In order to isolate the quantum corrections in which we are interested, exploiting
the fact that $\sigma$ is known with very high precision, we define the new 
observable
\beq
\label{Rwp}
\Rw^{'}(L,\frac{L}{u},n)=
\Rw(L,\frac{L}{u},n)-
\exp \left\{-\sigma n\left(n+L(1-\frac{1}{u})\right) \right\}~.
\eeq

\begin{figure}
\centering
\begin{tabular}{cc}
\begin{minipage}{200pt}
\includegraphics[width=1.\textwidth]{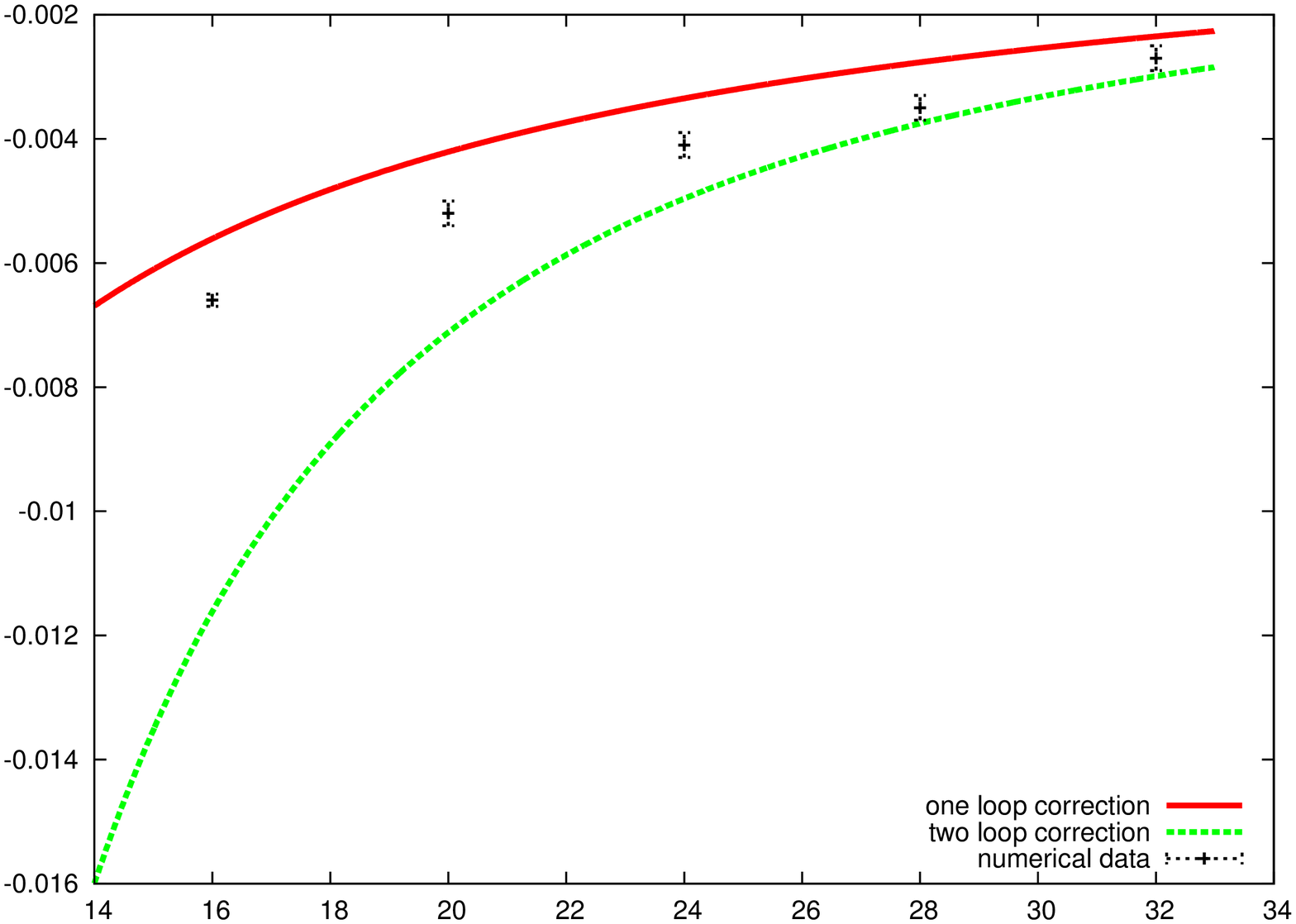}
\caption*{$  \Rw^{'}(L,\frac{3}{4}L,1) $}
\end{minipage} &
\begin{minipage}{200pt}
\includegraphics[width=1.\textwidth]{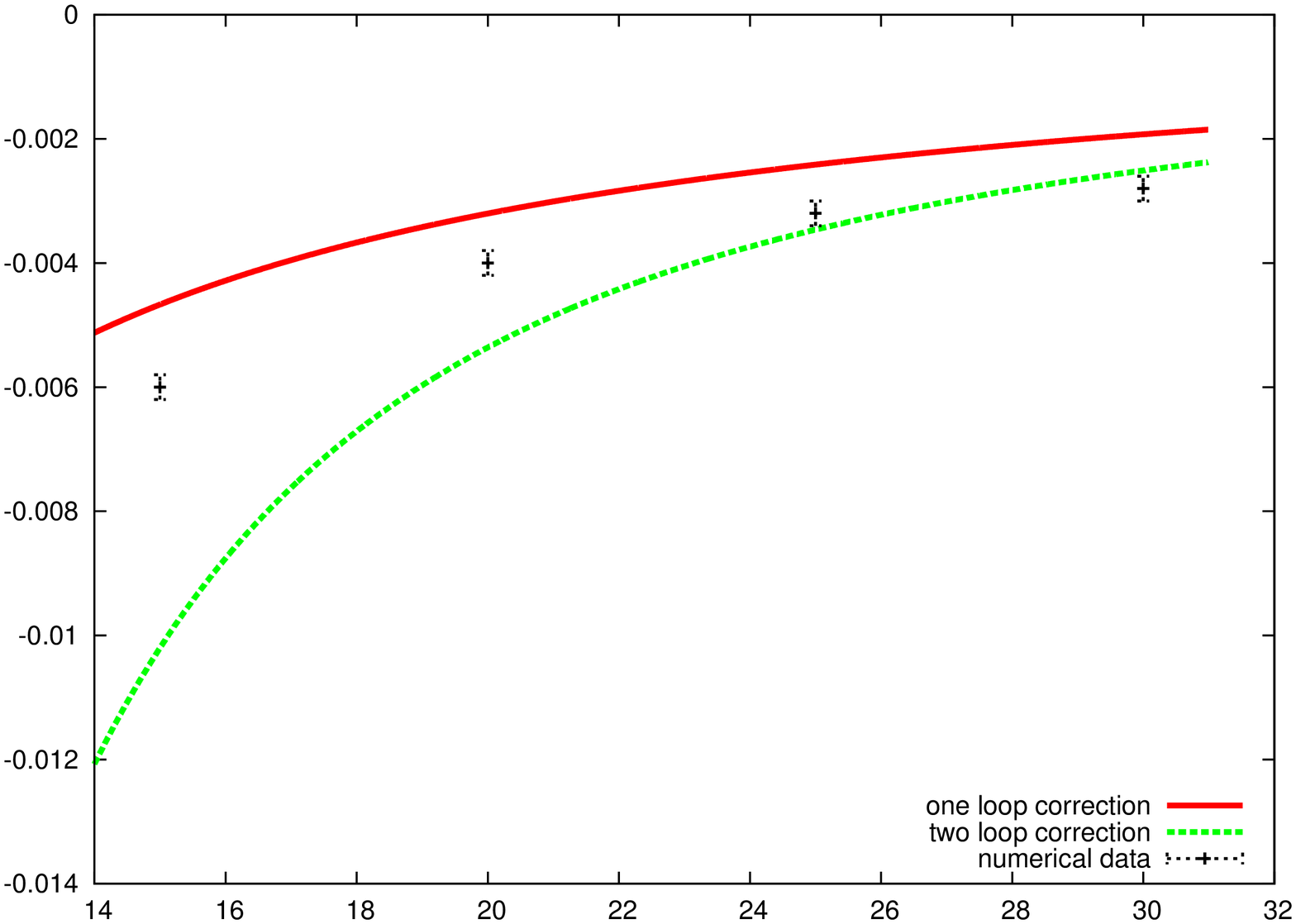}
\caption*{$ \Rw^{'}(L,\frac{4}{5}L,1) $}
\end{minipage} \\
\begin{minipage}{200pt}
\includegraphics[width=1.\textwidth]{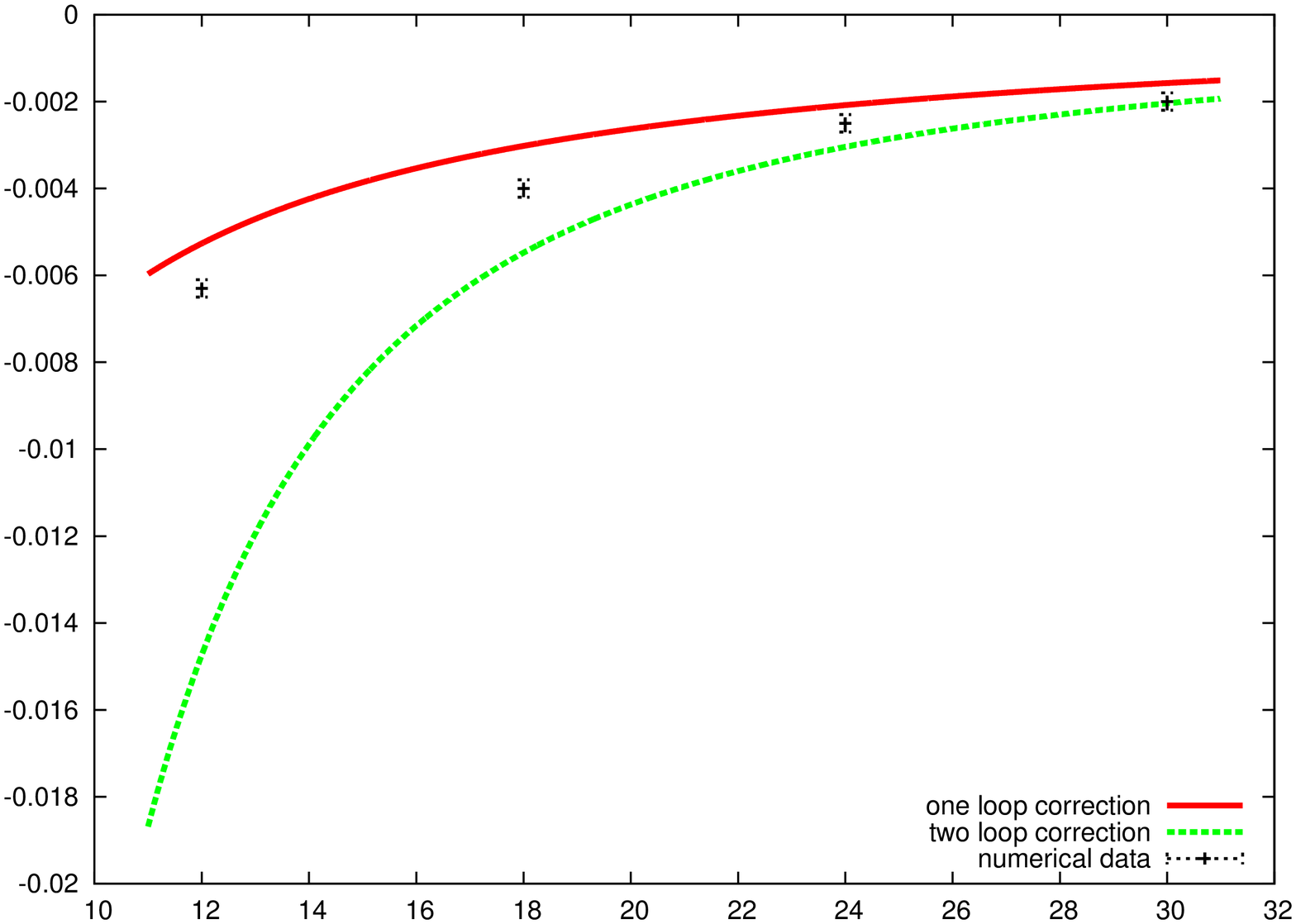}
\caption*{$  \Rw^{'}(L,\frac{5}{6}L,1) $}
\end{minipage} &
\begin{minipage}{200pt}
\includegraphics[width=1.\textwidth]{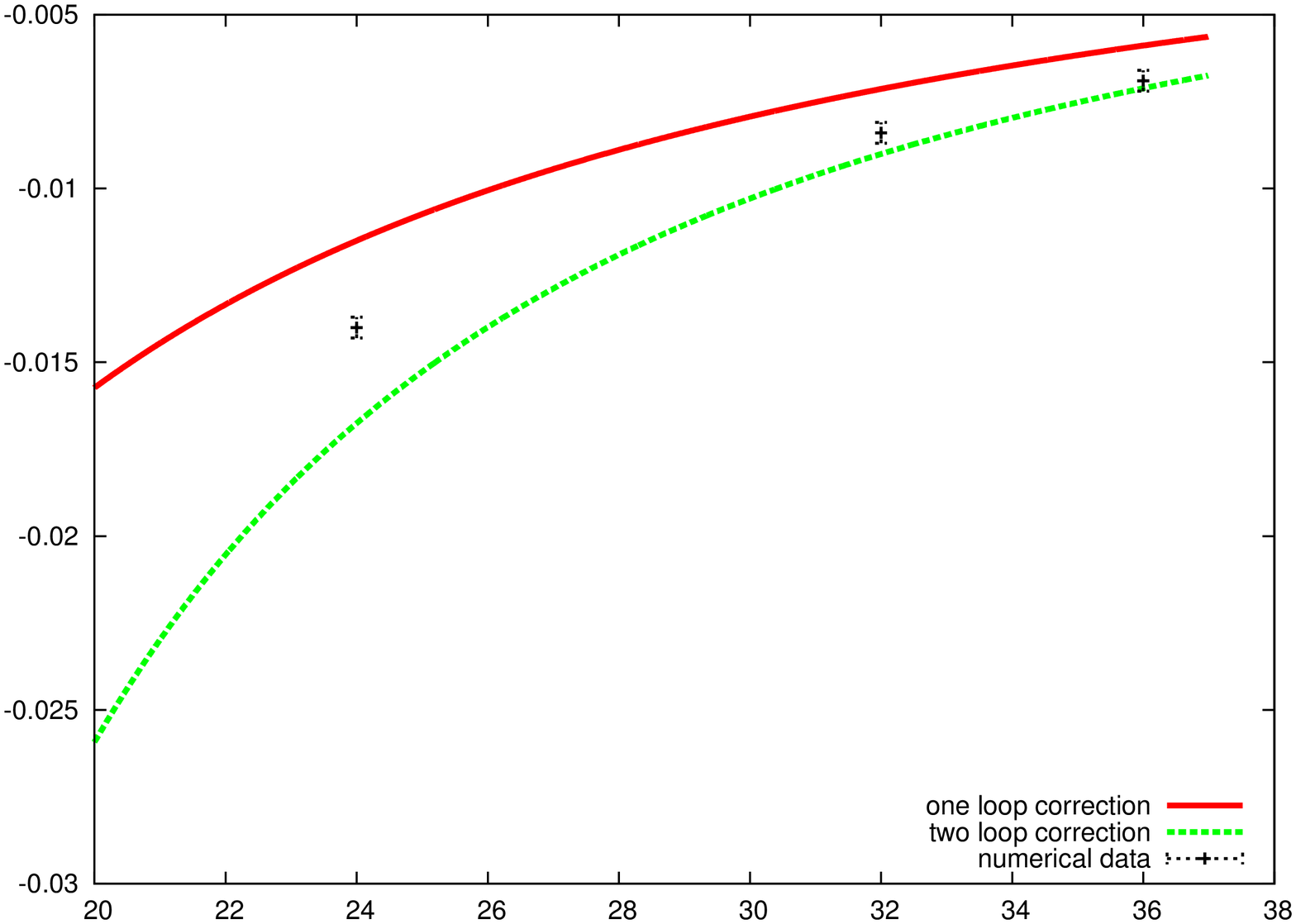}
\caption*{$  \Rw^{'}(L,\frac{3}{4}L,3) $}
\end{minipage}
\end{tabular}
\caption{Plot of $ \Rw^{'}(L,L/u,n) $ 
for different values of $u$ and $n$ against the quantum string corrections, 
up to two loop. Note that the classical 
string prediction is $\Rw^{'}=0$, see eq.s (\ref{Rwcl},\ref{Rwp}). }
\label{fig:results}
\end{figure}

We plot in fig. (\ref{fig:results}) our data for $ \Rw^{'} $ against
the quantum string corrections up to two loop.
Looking at these figures we see that a few interesting facts emerge.
\begin{enumerate} 
\item The deviation from the classical behavior $ \Rw^{'}=0 $ is immediately
visible from the plots: string corrections do definitely affect the Wilson loop
ratio.
\item The deviation of the data is systematically larger than the one loop prediction 
(which is indicated by the full line in fig.\ref{fig:results}).
For large enough values of $R$, the smallest of the two sides of the Wilson loop, 
the data converge toward the two loop string correction
(the dashed line in fig.\ref{fig:results}) and for the largest values of $L$ we 
considered ($R > 20$) they agree with it within the errors. 
\item This agreement becomes worse as $ R $ decreases. This also happens for
values of $ R $ much larger than the flux tube thickness and thus cannot be due
to the breaking of the effective string picture. The most probable reason 
is the presence of subleading effects due to boundary terms in the effective string action. 
\end{enumerate}
The first point is a well established result: effective string corrections
for the Wilson loop were observed for the first time in the gauge Ising model in
\cite{Caselle:1996ii} and later also in several other LGTs and in particular in
the - physically relevant - four dimensional SU(3) LGT \cite{Necco:2001xg}.
The second point, instead, is a new result in the Wilson loop geometry.   

Let us address in more detail the third point. At present there is no explicit
evaluation for the quantum effects due to boundary terms in the Wilson loop
geometry. This is a very interesting and non trivial issue which we plan to
address in future work. However, extrapolating to the Wilson case the existing
results~\cite{Aharony:2010db,Brandt:2010bw} for the cylinder geometry we expect
that these corrections should decrease at least as $1/R^4$. This behaviour seems
indeed to fit rather well the deviations observed in fig.\ref{fig:results}.
Extrapolating the fits to larger values of $R$ one can show that for $R > 20$
boundary corrections are fully negligible. This statement holds true assuming
for the boundary corrections any possible decreasing behaviour of the type
$1/R^n$ with $n\geq4$, assuming any value, compatible with the data at low $R$,
for the amplitude of the correction and, most importantly, making no assumptions
on the effective string corrections, except for those implied by the universality
theorems. 

This makes us confident that our results for the largest values of $R$ can be
used for an unbiased test of our theoretical predictions. In this respect it is
interesting to observe that in this region our data suggest that the universal
three loop correction should be small or should have the opposite sign with
respect to the two loop one. In order to clarify this point, however, we need to
compute this correction. This is the main goal of the following section.
  
\begin{table}[ht]
\centering
\phantom{-------}
\begin{tabular}{|cc|cc|cc|cc|}
\hline
$ L $ & $\Rw(L,\frac{3}{4}L,1) $ & $ L $ & $ \Rw(L,\frac{4}{5}L,1) $ 
& $ L $ & $ \Rw(L,\frac{5}{6}L,1) $ & $ L $ & $\Rw(L,\frac{3}{4}L,3) $  \\
\hline
 16 & 0.9421(1) & 15 & 0.9528(2)  & 12 & 0.9626(1) & 24 & 0.7387(3) \\
 20 & 0.9336(2) & 20 & 0.9447(2)  & 18 & 0.9548(2) & 32 & 0.6982(3) \\
 24 & 0.9249(2) & 25 & 0.9356(2)  & 24 & 0.9462(2) & 36 & 0.6777(3) \\
 28 & 0.9158(2) & 30 & 0.9262(2)  & 30 & 0.9368(2) &  \ &     \     \\
 32 & 0.9069(2) &  \ &          \ &  \ &     \     &  \ &     \     \\
  \hline
\end{tabular}
\phantom{-------}
\caption{Results for $ \Rw^{'}(L,R,n) $ as a function of the long side of the loop, 
for various fixed values of the ratio $u=L/R$ and of $n$.}
\label{results}
\end{table}

\section{Beyond two-loops: operatorial approach} 
\label{sec:beyond} 
Having argued that it is desirable to compute the three loop correction to the
Wilson loop amplitude in the NG model, we proceed now to do so by resorting to an
approach based on the first order re-formulation of the model itself. This
approach has been used in \cite{Billo:2005iv} and \cite{Billo:2006zg} for the
cylinder and torus partition functions respectively. The reasons to take this
route are two-fold. On the one hand, carrying out the derivation of the three
loop effect in the physical gauge by extending the two-loop computations of
\cite{Dietz:1982uc} is a daunting task. On the other hand, the application of
the operatorial, first-order formalism to the Wilson loop disk partition
function is not difficult but also not trivial and, as we will see, 
relies on the use of the rather non-standard (but perfectly sensible) concept of
open string boundary states introduced in \cite{Imamura:2005zm}. The way in
which the loop expansion of the Wilson loop amplitude emerges in this formalism
is quite interesting.  

We use the first order formulation, requiring the introduction of an independent
world-sheet metric $\gamma_{ab}$. This metric can then be gauge-fixed exploiting 
reparametrization invariance, and the open string action in the conformal gauge
\footnote{In the
conformal gauge
the world-sheet metric is of the form $\gamma_{\alpha\beta} = \ee^\phi
\delta_{\alpha\beta}$, and corresponds to a CFT of central charge
$c_{\mathrm{gh.}} = -26$. The scale factor $\ee^\phi$ decouples at the classical
level, but this property persists at the quantum level only if the anomaly
parametrized by the total central charge $c= D - 24$ vanishes. We will
nevertheless proceed in the case of general $D$, according to the discussion in
the introduction.} 
reads simply
\beq
\label{sac}
S = \sigma\int d\xi^1 \int_0^\pi d\xi^2\,
\left[\left(\partial_{\xi^1} X^M\right)^2 + 
\left(\partial_{\xi^2} X^M\right)^2
\right] + S_{\mathrm{gh.}}~,
\eeq
where $\xi^2\in [0,\pi]$ parametrizes the spatial extension of the string and
$\xi^1$ its proper (euclidean) time evolution.
The fields $X^M(\xi^1,\xi^2)$, with $M=1,\ldots,D+2$, describe the embedding of
the string world-sheet in the euclidean target space and form the 2-dimensional CFT of
$D+2$ free bosons. The term $S_{\mathrm{gh.}}$ in \eq{sac} is the action for the
ghost and anti-ghost fields
% (traditionally called $c$ and $b$) 
that arise from the Jacobian for fixing the conformal gauge. We do not really need here its
explicit expression, see \cite{Green:1987sp} or \cite{Polchinski:1998rq} for
reviews. 

We can treat this theory in an ``Hamiltonian'' way, by selecting the coordinate
$\xi^1$ as a (radial, euclidean) ``time'' coordinate. At fixed $\xi^1$, the
fields $X^M$ describe the embedding of the string in the target space; evolving
in $\xi^1$ the string sweeps out a surface. 

\subsection{The Dirichlet string}
\label{subsec:dir}
\begin{figure}
\def\svgwidth{14.5cm}
\begin{center}
\begingroup
  \makeatletter
  \providecommand\color[2][]{%
   \errmessage{(Inkscape) Color is used for the text in Inkscape, but the package 'color.sty' is not loaded}
    \renewcommand\color[2][]{}%
  }
  \providecommand\transparent[1]{%
    \errmessage{(Inkscape) Transparency is used (non-zero) for the text in Inkscape, but the package 'transparent.sty' is not loaded}
    \renewcommand\transparent[1]{}%
  }
  \providecommand\rotatebox[2]{#2}
  \ifx\svgwidth\undefined
    \setlength{\unitlength}{601.6015625pt}
  \else
    \setlength{\unitlength}{\svgwidth}
  \fi
  \global\let\svgwidth\undefined
  \makeatother
  \begin{picture}(1,0.3525291)%
    \put(0,0){\includegraphics[width=\unitlength]{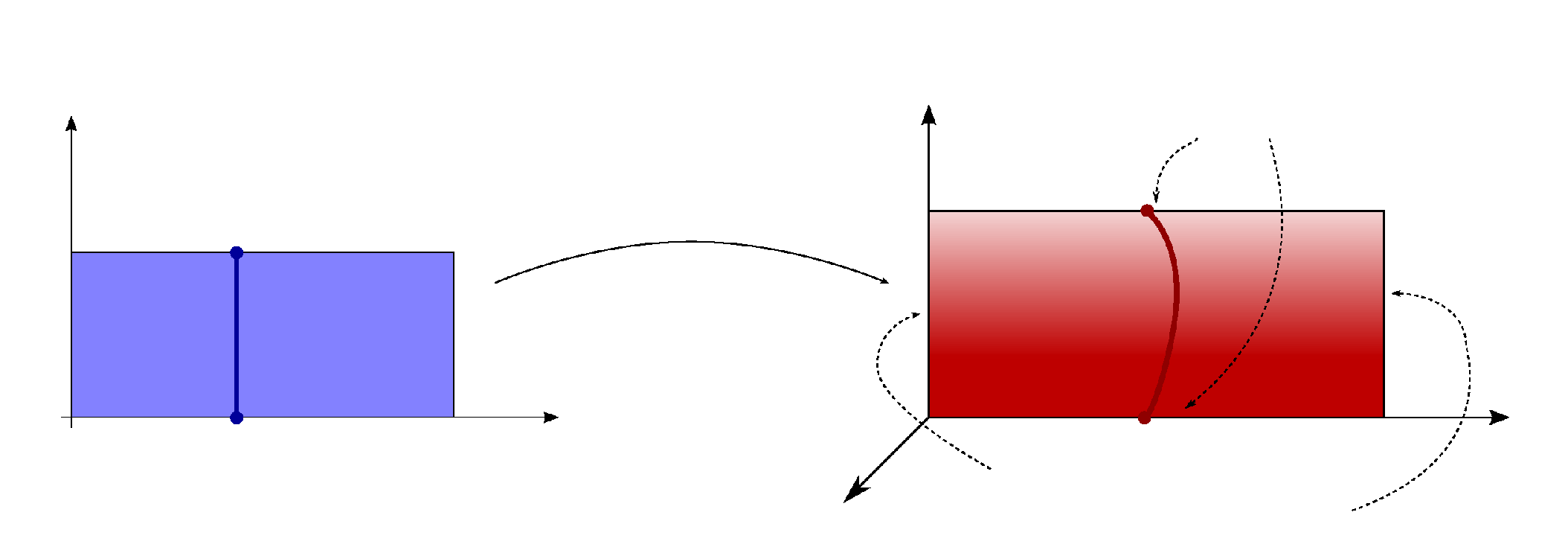}}%
    \put(0.54554462,0.26060987){\makebox(0,0)[lb]{\smash{$x^2$}}}%
    \put(0.52048112,0.05350626){\makebox(0,0)[lb]{\smash{$\vec x$}}}%
    \put(0.93847103,0.04751373){\makebox(0,0)[lb]{\smash{$x^1$}}}%
    \put(0.66619533,0.29901532){\makebox(0,0)[lb]{\smash{{\tiny open string with b.c.'s}}}}
    \put(0.66672994,0.27550074){\makebox(0,0)[lb]{\smash{{\tiny as in eq.s (\ref{nbcp}-\ref{condtran})}
}}}%
    \put(0.33448361,0.05380782){\makebox(0,0)[lb]{\smash{$\xi^1$}}}%
    \put(-0.00084636,0.27337281){\makebox(0,0)[lb]{\smash{$\xi^2$}}}%
    \put(0.00029802,0.19206299){\makebox(0,0)[lb]{\smash{$\pi$}}}%
    \put(0.27776097,0.05266344){\makebox(0,0)[lb]{\smash{$t$}}}%
    \put(0.11155043,0.33052383){\makebox(0,0)[lb]{\smash{\emph{World-sheet}}}}%
    \put(0.69856385,0.33524255){\makebox(0,0)[lb]{\smash{\emph{Target space}}}}%
    \put(0.39648278,0.21180201){\makebox(0,0)[lb]{\smash{$X^M(\xi^1,\xi^2)$}}}%
    \put(0.62390069,0.03085832){\makebox(0,0)[lb]{\smash{{\tiny Emitted}
}}}%
    \put(0.82265019,0.00333132){\makebox(0,0)[lb]{\smash{{\tiny Readsorbed}
}}}%
    \put(0.86701524,0.04751373){\makebox(0,0)[lb]{\smash{$L$}}}%
  \end{picture}%
\endgroup
\end{center}
\caption{A surface bordered by the rectangular Wilson loop, such as the one drawn
on the left, can be spanned by an open string with end-points attached to the sides at $x^2$ 
and $\vec x$ fixed but free to move in the $x^1$ direction. This happens if such a string is emitted 
from the vacuum at $x^1=0$ and re-adsorbed at $x^1=L$.}
 \label{fig:setup}
\end{figure}

We argue that such an operatorial description%
\footnote{Our operatorial approach differs from the outset from the
path-integral approach to the Polyakov model applied to the Wilson loop topology
in \cite{Alvarez:1982zi,NBI-HE-83-21}. In that case, Dirichlet boundary
conditions (modified so as to take into account reparametrization invariance
along the boundary) are imposed along the entire Wilson  loop contour.} is 
possible also with the boundary conditions corresponding  to a rectangular
Wilson loop, as described in Fig. \ref{fig:setup}. We
arbitrarily  select one of the directions along the Wilson loop, say $x^1$, and
regard the sides of the loop along which $x^1$ varies, which sit at distance $R$
in the $x^2$ direction, as the quark and anti-quark lines. Consider a open
string whose end-points are attached to these two lines and are free to move in
the $x^1$ direction. If such a string is emitted from the vacuum at the Wilson
loop side placed at $x^2=0$ and re-adsorbed at the side placed at $x^2=L$ then
it spans a surface bordered by the Wilson loop. 

In this description the directions $x^1$ and $x^2$ play a very
asymmetric r\^ole; we must however ensure that the Wilson loop partition
function we compute be invariant under the exchange of $L$ and $R$. 
To make this picture concrete, one has to quantize the open string with the boundary 
conditions just discussed, and construct then in its Hilbert space
the states which  describe its emission from the vacuum. Such states
represent the open string analogue of
the boundary states that describe the insertion of a boundary in the closed string world-sheet, i.e., 
the emission of closed strings from the vacuum (or from D-branes)
\cite{PUPT-1070}, see for instance \cite{DiVecchia:1999rh} for a review.

Let us start by imposing on the string the boundary conditions discussed above.
These are of  Neumann typs conditions at both ends in the
$x^1$ direction:
\beq
\label{nbcp}
\left.\partial_{\xi^2} X^1(\xi^1,\xi^2)\right|_{\xi^2=0,\pi} = 0~,
\eeq
and of Dirichlet type at both ends in the direction
$x^2$: 
\begin{equation}
\label{x1bc}
X^2(\xi^1,0) = 0~,~~~ X^2(\xi^1,\pi) = R~,
\end{equation}
and in the transverse ones:
\begin{equation}
\label{condtran}
\vec X(\xi^1,0) = \vec X(\xi^1,\pi) = \vec 0~.
\end{equation}

With these boundary conditions, the string fields admit the following
expansions:
\begin{equation}
\label{expx0}
\begin{aligned}
X^1(\xi^1,\xi^2) & = \hat x^1 + \frac{\sigma}{2} \hat p^1\xi^1 + \ii 
\sqrt{\frac{\sigma}{2}} 
\sum_{n=1}^\infty \left(\frac{a^1_n}{\sqrt{n}}\ee^{-n \xi^1} 
- \frac{(a^1_n)^\dagger}{\sqrt{n}}\ee^{n \xi^1}\right) \cos n\xi^2~,
\\
X^2(\xi^1,\xi^2) & = \frac{R}{\pi} \xi^2 - 
\sqrt{\frac{\sigma}{2}} 
\sum_{n=1}^\infty \left(\frac{a^2_n}{\sqrt{n}}\ee^{-n \xi^1} 
+ \frac{(a^2_n)^\dagger}{\sqrt{n}}\ee^{n \xi^1}\right)
\sin n\xi^2~,\\
\vec X(\xi^1,\xi^2) & =  - 
\sqrt{\frac{\sigma}{2}} 
\sum_{n=1}^\infty \left(\frac{{\vec a}_n}{\sqrt{n}}\ee^{-n \xi^1} 
+ \frac{(\vec{a}_n)^\dagger}{\sqrt{n}}\ee^{n \xi^1}\right)\sin n\xi^2~.
\end{aligned}
\end{equation}

Canonical quantization leads to the following commutation relations among the
oscillators $a^M_n$: 
\beq
\label{osc_open}
\comm{a^M_m}{{(a^N_n)^\dagger}} = \delta_{m,n}\,\delta^{MN}~.
\eeq

The modes of the stress-energy tensor, usually denoted as $L_m$, generate the
residual conformal transformations of the world-sheet. In particular, $L_0$
generates the world-sheet dilations and corresponds to the Hamiltonian derived
from the action \eq{sac}. It receives contributions from the bosons and the
ghost system: $L_0= L_0^{(X)} + L_0^{(\mathrm{gh.})}$, and we have
\beq
\label{L0Xo}
L_0^{(X)} = \frac{ (\hat p^1)^2}{2\pi\sigma} + 
\frac{\sigma}{2\pi} R^2+
\sum_{n=1}^\infty n\, N_n - \frac{d}{24}~,
\eeq
where $N_n = \sum_M a^M_{-n}\cdot a^M_{n}$ is the occupation number for
the oscillators $a^M_n$, and $d/24$ is the ($\zeta$ function regularized%
\footnote{In \cite{Orland:2001rq} some words of caution were
raised regarding the use of $\zeta$-function regularization for the string with
the present boundary conditions. However, in \cite{Billo:2005iv} it was shown
that in the Polyakov loop correlator geometry, where open strings have very
similar boundary conditions, the results of this procedure are perfectly
consistent.}) normal ordering constant. The term $\frac{\sigma}{\pi} R^2$
represents the energy due to the stretching of the string between the two
opposite sides of the Wilson loop. For the ghost system we have, see for
instance \cite{Green:1987sp},
\beq
\label{L0gh}
L_0^{\gh} = \mbox{non-zero modes} + \frac{1}{12}~. 
\eeq
In the computation of partition functions, the net effect of the non-zero mode
ghost modes (which have the opposite statistics with respect to ordinary bosonic
fields) is to cancel the non-zero-mode contribution of two bosonic fields, 
reconciling this treatment with the static gauge description where only the 
transverse fields are present to begin with. 
This happens when the operatorial approach is utilized to derive the 
effective string predictions for Polyakov loop correlators \cite{Billo:2005iv}
and for interfaces \cite{Billo:2006zg} and 
this will be the case also for the Wilson loop. 

\subsection{The Wilson loop amplitude}
\label{subsec:wlp}
We have implemented the boundary conditions along the two sides in the direction 
$x^1$ of the Wilson
loop in the definition of the Dirichlet string. We have now to enforce the boundary
conditions at the two sides in the direction $x^2$ as operatorial condition on
suitable states in the Hilbert space of this string. We proceed by constructing 
``open string boundary states'' \cite{Imamura:2005zm}, which we denote as
$\ket{\Bop,0}$ and $\ket{\Bop,L}$,  
 such that at proper time $\xi^1=0$ we have
\begin{align}
\label{b0}
X^1(0,\xi^2)\ket{\Bop,0} & = 0~, &  X^1(0,\xi^2)\ket{\Bop,L} & = L
\ket{\Bop,L}~,\\
\notag
X^2(0,\xi^2)\ket{\Bop,0} & = \frac{R}{\pi}\xi^2\,\ket{\Bop,0}~,& 
X^2(0,\xi^2)\ket{\Bop,L} & = \frac{R}{\pi}\xi^2\,\ket{\Bop,L}~,\\
\notag
\vec X(0,\xi^2)\ket{\Bop,0} & = \vec 0~, & \vec X(0,\xi^2)\ket{\Bop,L} & = \vec
0~.
\end{align}
The conditions on $\ket{\Bop,0}$ mean that, when applied to this state,
the string fields $X^M(\xi^2)$ describe an open string coinciding with  
the side of the Wilson loop at $x^1=0$. Analogous is the meaning of the conditions
on $\ket{\Bop,L}$.

We propose that the following expression:
\begin{equation}
\label{amplin1}
\cW(L,R) = \int_0^\infty \frac{dt}{t^\omega}\, \bra{\Bop,0}\ee^{-2\pi t L_0 }
\ket{\Bop,L}~,
\end{equation}
which we will call ''Wilson loop amplitude'', be proportional to the Wilson loop
free energy $W(L,R)$. This expression is obtained by integrating over the real
parameter $t$ the quantum mechanical kernel describing the propagation of the
string from the state $\ket{\Bop,0}$ to the state $\ket{\Bop,0}$ in a proper
Euclidean time time $2\pi t$. If $\omega$ were equal to zero, \eq{amplin1} would
contain the open string propagator $L_0^{-1}$ (written via the Schwinger
parametrization), sandwiched between two boundary states. This is the same kind
of expression used to re-formulate the open string cylinder partition function
as the tree level propagation of a closed string between two boundary states
\cite{PUPT-1070,DiVecchia:1999rh} that was applied to the effective description
of Polyakov loop correlators in \cite{Billo:2005iv}.

However, in order to obtain an expression invariant under the exchange
$L\leftrightarrow R$ we will be forced to choose a non-zero value of $\omega$,
see later \eq{omechoice}. We do not have an a priori understanding of this
choice; if we accepted, however, this choice not only guarantees the
$L\leftrightarrow R$ symmetry, but reproduces the results of the static gauge
approach at the classical, one-loop and two-loop level, which represents a very
non trivial check. This a posteriori confirmation makes us confident that we can
use \eq{amplin1} to derive higher loop contributions; this can be done with a
very limited effort, compared to the static gauge approach. Indeed, $\cW(L,R)$
can be expressed as an infinite sum of Bessel function contributions; moreover,
it can be rather straightforwardly expanded in inverse powers of $1/(\sigma
LR)$, i.e., in sigma-model loops of the NG formulation. To see this, we now
consider the various ingredients of the amplitude \eq{amplin1}.

\paragraph{Open string boundary states}
The open string boundary states have the form 
\begin{equation}
\label{bsdec}
\ket{\Bop,0} = \mathcal{N}\, \ket{\Bop,0}_{(0)} \, 
\ket{\Bop,0}_{\mathrm{(n.z.)}}\, \ket{\Bop,0}_{\mathrm{(gh.)}}~;
\end{equation}
of, course, the same decomposition applies to $\ket{\Bop,L}$. Here,
$\mathcal{N}$ is a normalization which will be irrelevant for our purposes, 
while the subscripts refer to the zero-mode,
non-zero-mode and ghost sectors of the open string Hilbert space. Let us
analyze in turn these various components.
From the expansion \eq{expx0}  of the string fields it follows that the
conditions \eq{b0} in the zero-mode sector only imply that the c.o.m. position
operator $\hat x^1$ assumes the values $0$ and $L$ on $\ket{\Bop,0}$ and
$\ket{\Bop,L}$ respectively. The zero-mode part of these states is therefore
given by 
\begin{align}
\label{zmb}
\ket{\Bop,0}_{\mathrm{(0)}} & = \delta({\hat x}^1)\ket{0} = \frac{1}{2\pi}\int
dq_1 \ket{q_1}~,\\
\ket{\Bop,L}_{\mathrm{(0)}} & = \delta({\hat x}^1-L)\ket{0} = 
\frac{1}{2\pi}\int dq^\prime_1 \ee^{-\ii q_1^\prime L}\ket{q_1^\prime}~,
\end{align}
where $\ket{0}$ is the vacuum in the zero-mode sector, on which all components
$\hat p^M$ of the momentum operator vanish, while $\ket{q_1}$ is an eigenstate
of ${\hat p}_1$ with eigenvalue $q_1$ (the other components of the momentum
vanish).  

The non-zero mode sector of the Hilbert space corresponds to a collection of
harmonic 
oscillators. The conditions \eq{b0} imply, through the mode expansions
\eq{expx0}, that, for any $n$, 
\begin{equation}
\label{condnz}
\left(a^1_n + a^{1\dagger}_n\right) \ket{\Bop,0}_{\mathrm{(n.z.)}} = 0~,~~
\left(a^2_n - a^{2\dagger}_n\right) \ket{\Bop,0}_{\mathrm{(n.z.)}} = 0~,~~
\left({\vec a}_n - \vec{a}_n^\dagger\right) \ket{\Bop,0}_{\mathrm{(n.z.)}} = 0~;
\end{equation}
exactly the same holds for  $\ket{\Bop,L}_{\mathrm{n.z.}}$. 
These conditions are easily solved%
\footnote{For an harmonic oscillator of mass $m$ and frequency $\omega$, 
such type of conditions correspond to the vanishing of either
the momentum $\hat p = \sqrt{\frac{m\hbar\omega}{2}} (a + a^{\dagger})$ or the position 
$\hat x = \ii \sqrt{\frac{\hbar}{2m\omega}} (a - a^{\dagger})$ operator.
Momentum or position eigenstates can be expressed in terms of the vacuum as 
\begin{equation}
\label{posmomosc}
\ket{\hat x = x} = \left(\frac{m\omega}{\hbar\pi}\right)^{\frac 14}\ee^{\frac 12 (a^\dagger)^2 - \ii
\sqrt{\frac{2m\omega}{\hbar}}x a^\dagger - \frac 12 \frac{m\omega}{\hbar} x^2}\ket 0~,~~~
\ket{\hat p = p} =  \left(\frac{\hbar}{\pi m\omega}\right)^{\frac 14}\ee^{-\frac 12 (a^\dagger)^2 + \sqrt{\frac{2}{m\hbar\omega}}p
a^\dagger -\frac{p^2}{2\hbar\omega m}}\ket 0~.
\end{equation}
}
and we have 
\begin{equation}
\label{solbnz}
\ket{\Bop,0}_{\mathrm{n.z.}} = \ket{\Bop,L}_{\mathrm{n.z.}}=
\bigotimes_{n=1}^\infty \ee^{\frac 12 \left(- (a^{1\dagger}_n)^2 +
(a^{2\dagger}_n)^2
+ ({\vec a}^\dagger_n)^2\right)}{\ket 0}_{(n)}~,
\end{equation}  
where by ${\ket 0}_{(n)}$ we denote the tensor vacua for the $n$-th oscillation
modes in all directions, and the overall normalization was already included in \eq{bsdec}.

To avoid excessive technicality, we do not write explicitly the form of the boundary state
in the ghost sector

\paragraph{The amplitude}
We can now compute the matrix element appearing in the definition \eq{amplin1} of
the Wilson amplitude.
From the explicit expression of the Virasoro generator $L_0$, \eq{L0Xo}, in the
zero-mode sector we find
\begin{equation}
\label{zmmatel}
\begin{aligned}
{}_{(0)}\bra{\Bop,0}\ee^{-\pi t L_0 } \ket{\Bop,L}_{(0)} & =
\int \frac{dq_1}{2\pi} \frac{dq_1^\prime}{2\pi} \bra{q_1} 
\ee^{-2\pi t\left(\frac{({\hat p}_1)^2}{2\pi\sigma} + \frac{\sigma}{2\pi}
R^2\right)} 
\ee^{-\ii q_1^\prime L}\ket{q_1^\prime}
\\
& = \sqrt{\frac{2\pi\sigma}{t}} \ee^{-\sigma R^2 t - \frac{\sigma\, L^2}{4 t}}~,
\end{aligned}
\end{equation}
where we used the orthogonality of momentum eigenstates:
$\braket{q_1}{q_1^\prime} = 
2\pi\delta(q_1-q_1^\prime)$ and performed the remaining Gaussian integration.

In the non-zero-mode sector, relying on the oscillator algebra it is not
difficult to compute%
\footnote{Notice that all the $D+2$ direction contribute in the same way to the
matrix element, despite the fact that  the exponent in the boundary state
\ref{solbnz} has the opposite sign in front of the $(a^{1\dagger}_n)^2$
oscillators with respect to the other ones.}
\begin{equation}
\label{nzmatel}
\begin{aligned}
& {}_{(\mathrm{n.z.})}\bra{\Bop,0}\ee^{-2\pi t L_0 } 
\ket{\Bop,L}_{(\mathrm{n.z.})}\\
= & \ee^{\pi t\frac{D+2}{24}}\,
\prod_{n=1}^\infty 
{}_{(n)}\bra{0}  \ee^{\frac 12 \left(- (a^{1}_n)^2 + (a^{2}_n)^2
+ ({\vec a}_n)^2\right)} \ee^{-2\pi t\, n N_n} 
 \ee^{\frac 12 \left(- (a^{1\dagger}_n)^2 + (a^{2\dagger}_n)^2
+ ({\vec a}^\dagger_n)^2\right)} \ket{0}_{(n)} \\
= &  q^{-\frac{D+2}{48}}\,\prod_{n=1}^\infty (1 - q^{n})^{-\frac{D+2}{2}} =
  \left[\eta(\tau)\right]^{-\frac{D+2}{2}}~,
\end{aligned}
\end{equation}
where we introduced, for later notational convenience,
\begin{equation}
\label{defq}
q = \ee^{-4\pi t} \equiv \ee^{2\pi\ii\tau}
\end{equation}
and wrote the result in terms of the Dedekind eta function 
(see Appendix \ref{app:not} for our conventions). Recall that $t$ is a real parameter.
 
The matrix element in the ghost sector is such that it cancels exactly the
non-zero-mode contribution of two directions. Thus, putting together all the pieces, 
we obtain 
%the following expression for the Wilson loop:
\begin{equation} 
\label{wle1}
\cW(L,R) = \sqrt{2\pi\sigma} |\mathcal{N}|^2 \int_0^\infty \frac{dt}{t^{\omega
+
\frac 12}} 
\ee^{-\sigma R^2 t - \frac{\sigma\, L^2}{4 t}}\,
\left[\eta(\tau)\right]^{-\frac{D}{2}}~.
\end{equation}
Let us consider the expression $\cW(R,L)$ obtained exchanging $L$ and $R$. Let
us change integration variable in this expression by setting 
$t = 1/(4t^\prime)$,
which corresponds, according to \eq{defq}, to $\tau = -1/\tau^\prime$. We can
then use the modular properties of Dedekind's function, see \eq{mteta},
and find 
\begin{equation} 
\label{wlex}
\cW(R,L) = \sqrt{2\pi\sigma} |\mathcal{N}|^2 
2^{2\omega -1 - \frac {d-2}{4}}\,
\int_0^\infty 
\frac{d t^\prime}{{t^\prime}^{\frac 32 -\omega + 
\frac {d-2}{4}}}\, 
\ee^{-\sigma R^2 t^\prime - \frac{\sigma\, L^2}{4 t^\prime}}\, 
\left[\eta(\tau^\prime)\right]^{-\frac{D}{2}}~.
\end{equation}
%Renaming again $t$ the integration variable, 
This coincides with $\cW(L,R)$, as given in \eq{wle1}, provided we fix 
\begin{equation}
\label{omechoice}
\omega = \frac{D}{8} + \frac 12~.
\end{equation} 

\subsection{Explicit integration and loop expansion}
\label{subsec:explint}
The integral over the parameter $t$ in the definition of the Wilson loop
amplitude can be explicitly performed, by techniques very similar to those
exploited in \cite{Billo:2005iv,Billo:2006zg} to handle other geometries in the
same operatorial approach to the effective string. We can Fourier expand the
power of Dedekind's eta function appearing in \eq{wlex} writing
\begin{equation}
\label{etaexp}
\left[\eta(\tau)\right]^{-\frac{D}{2}} = 
\sum_{k=0}^\infty c_k \ee^{-4\pi t \left(k - \frac{D}{48}\right)}
= \sum_{k=0}^\infty c_k\, q^{k - \frac{D}{48}}
= \sum_{k=0}^\infty c_k\, q^{\hat k}~,
\end{equation} 
where for simplicity we introduced
\begin{equation}
\label{defkhat}
\hat k = k - \frac{D}{48}~.
\end{equation}
Taking into account \eq{omechoice} we have then 
\begin{equation}
\label{wlex2}
\cW(L,R)  = \sqrt{2\pi\sigma} |\mathcal{N}|^2 \sum_{k=0}^\infty c_k\, 
\int_0^\infty \frac{dt}{t} t^{-\frac{D}{8}} 
\ee^{-\sigma\left[\frac{L^2}{4t} + R^2 t + \frac{4\pi t}{\sigma}\hat k\right]}~.
\end{equation}    
We can now express the integral in terms of modified Bessel functions%
\footnote{We use the formula
\begin{equation}
\label{intbess}
\int_0^\infty \frac{dt}{t} t^{-\gamma} \ee^{-A^2 t - \frac{B^2}{t}} = 
2 \left(\frac{A}{B}\right)^\gamma \, K_\gamma(2 A B)~.
\end{equation}
} 
and write 
\begin{equation}
\label{wlexn2}
\cW(L,R)  = 2 \sqrt{2\pi\sigma} |\mathcal{N}|^2 \sum_{k=0}^\infty c_k\, 
\left(\frac{2 \cale_k}{u}\right)^{\frac{\alpha}{4}} \,
K_{\frac{\alpha}{4}}(\sigma \cA\, \cale_k)
~,
\end{equation}    
where for brevity we set
\begin{equation}
\label{gammad}
\alpha = \frac{D}{2}
\end{equation}
and introduced
\begin{equation}
\label{Ekdef}
\cale_k =  \sqrt{1 + \frac{4\pi u}{\sigma \cA}\hat k}~.
\end{equation}    

Eq. (\ref{wlex2}) is well suited for an expansion for large area (measured in
units of the inverse string tension), i.e., for large $\sigma \cA$, keeping
finite the ratio $u = L/R$. As we saw in section \ref{sec:twoloop}, in the physical
gauge Nambu-Goto description this corresponds to the loop expansion of the
sigma-model. Using the asymptotic expansion of the modified Bessel functions for
large values of their arguments, see \eq{asbessel}, and the expansion of $\cale_k$
we find 
\begin{equation}
\label{wlarge}
\begin{aligned}
\cW(\cA,u) & =  \sqrt{2\pi\sigma} |\mathcal{N}|^2
\left(\frac 2u\right)^\frac{\alpha}{4}\, \sqrt{\frac{\pi}{2\sigma\cA}}
\ee^{-\sigma\cA}
\sum_k c_k \ee^{-2\pi u \hat k}\\
& \times \Biggl\{ 1 + \frac{1}{\sigma\cA}\Bigl[
\frac{\alpha^2 - 4}{32} + \frac{\alpha-2}{2}\pi u \hat k + 2 \pi^2 u^2 
{\hat k}^2\Bigr]\\
& 
%\phantom{\times \Bigl\{ 1~\,} 
+ \frac{1}{(\sigma\cA)^2}\Bigl[
\frac{\alpha^4 - 40 \alpha^2 + 144}{2048} + \frac{\alpha^3 - 6\alpha^2 - 4\alpha
+ 24}{64}
\pi u \hat k   \\
& 
%\phantom{\times \Bigl\{ 1 + \frac{1}{(\sigma\cA)^2}\Bigl[}
+ \frac{3}{16}(\alpha^2 - 8\alpha + 12)
\pi^2 u^2 {\hat k}^2
+ (\alpha - 6) \pi^3 u^3 {\hat k}^3
%\\
%& 
%\phantom{\times \Bigl\{ 1 + \frac{1}{(\sigma\cA)^2}\Bigl[}
\, + 2  \pi^4 u^4 {\hat k}^4\Bigr]
%\\
%& 
%\phantom{\times \Bigl\{ 1~\,}
+ O\left(\frac{1}{(\sigma\cA)^3}\right)\Biggr\}
\end{aligned}
\end{equation}

Note that, recalling \eq{etaexp}, we have
\begin{equation}
\label{etau}
\sum_{k=0}^\infty c_k\, \ee^{-2\pi u \hat k} = \sum_k Q^{\hat k} =
\left[\eta(\ii u)\right]^{-\alpha}~;
\end{equation}
the introduction of $Q = \ee^{-2\pi u}$ is  convenient for later manipulations.
The powers of $\hat k$ appearing in 
\eq{wlarge} correspond then to powers of the logarithmic derivative $Q (d/dQ)$
applied to the sum in \eq{etau}. In other words, we have
\begin{equation}
\label{wlarge2}
\begin{aligned}
\cW(\cA,u) & =  \sqrt{2\pi\sigma} |\mathcal{N}|^2
\left(\frac 2u\right)^\frac{\alpha}{4}\, \sqrt{\frac{\pi}{2\sigma\cA}}
\ee^{-\sigma\cA}
\\
& \times \Biggl\{ 1 + \frac{1}{\sigma\cA}\Bigl[
\frac{\alpha^2 - 4}{32} + \frac{\alpha-2}{2}\pi u
\left(Q\frac{d\phantom{Q}}{dQ}\right) + 2 \pi^2 u^2 
\left(Q\frac{d\phantom{Q}}{dQ}\right)^2\Bigr]\\
& 
%\phantom{\times \Bigl\{ 1~\,} 
+ \frac{1}{(\sigma\cA)^2}\Bigl[
\frac{\alpha^4 - 40 \alpha^2 + 144}{2048} + \frac{\alpha^3 - 6\alpha^2 - 4\alpha
+ 24}{64}
\pi u \left(Q\frac{d\phantom{Q}}{dQ}\right)   \\
& 
%\phantom{\times \Bigl\{ 1 + \frac{1}{(\sigma\cA)^2}\Bigl[}
+ \frac{3(\alpha^2 - 8\alpha + 12)}{16}
\pi^2 u^2 \left(Q\frac{d\phantom{Q}}{dQ}\right)^2
+ (\alpha - 6) \pi^3 u^3\left(Q\frac{d\phantom{Q}}{dQ}\right)^3
%\\
%& 
%\phantom{\times \Bigl\{ 1 + \frac{1}{(\sigma\cA)^2}\Bigl[}
\, + 2  \pi^4 u^4 \left(Q\frac{d\phantom{Q}}{dQ}\right)^4\Bigr]
\\
& 
%\phantom{\times \Bigl\{ 1~\,}
+ O\left(\frac{1}{(\sigma\cA)^3}\right)\Biggr\}\,\left[\eta(\ii
u)\right]^{-\alpha}~.
\end{aligned}
\end{equation}
Logarithmic derivatives of the $\eta$ function can be expressed in terms of
Eisenstein series, see Appendix \ref{app:not} for details. By doing so,
recalling the expression \eq{gammad} of the parameter $\alpha$ in terms of the
dimension $D$, we finally get
\begin{equation}
\label{wlarge3}
%\begin{aligned}
\cW(\cA,u) 
%& 
= \sqrt{2\pi\sigma} |\mathcal{N}|^2
\, \sqrt{\frac{\pi}{2\sigma\cA}} \ee^{-\sigma\cA}\,\left(\frac
2u\right)^\frac{D}{8}\,
\left[\eta(\ii u)\right]^{-\frac{D}{2}}
%\\ & \times 
\Biggl\{ 1 + \frac{\hat \cL_2(u)}{\sigma\cA}
 + \frac{\hat \cL_3(u)}{(\sigma\cA)^2} + \ldots
\Biggr\}~,
%\end{aligned}
\end{equation}
where
\begin{equation}
 \label{2loopex}
\hat \cL_2(u) = 
\left(\frac{\pi}{24}\right)^2 \Bigl(2 D u^2 E_4(\ii u) -
\frac{D(D-4)}{2}
E_2(\ii u) E_2(\ii /u)\Bigr)
% \\ &
 + \frac{(D+4)(D-4)}{128}
\end{equation}
and
\begin{equation}
\label{3loopex}
\begin{aligned}
\hat \cL_3(u) & =  \left(\frac{\pi}{24}\right)^4  
\Bigl[6(D+24)D u^4 E_4^2(\ii u)
%\\ & 
- 3D(D-8)(D-12) u^2 E_4(\ii u) E_2(\ii u) E_2(\ii/u) \\
&  + \frac{D(D-4)(D-8)(D-12) }{8} E_2^2(\ii u) E_2^2(\ii/u) \Bigr]
\\ & 
- \left(\frac{\pi}{24}\right)^3 4 D(D-12) u^3 E_6(\ii u) 
\left(1 - \frac{\pi}{3} u E_2(\ii u)\right)
\\ & 
+ \left(\frac{\pi}{24}\right)^2\Bigl[\frac{3}{64}D(D-4)(D-12) u^2 E_4(\ii
u)\\
& -\frac{(D+4)D(D-4)(D-12)}{256} E_2(\ii u) E_2(\ii/u)\Bigr]
%\\ & 
+ \frac{(D+12)(D+4)(D-4)(D-12)}{32768}~.
\end{aligned}
\end{equation}

Using the transformation properties of the
Dedekind function and of the Eisenstein series given in Appendix \ref{app:not}
it is not difficult to check that \eq{wlarge3} is invariant
under the transformation $u \leftrightarrow 1/u$, i.e., under the exchange $L
\leftrightarrow R$.

The term in \eq{wlarge3} that multiplies the curly brackets should
correspond to the classical plus one-loop
result discussed in sec. (\ref{sec:twoloop}). 
%for the Wilson loop that appears as a prefactor in \eq{wlngexp}. 
%In terms of $L$ and $R$, and 
Up to numerical normalization, we have
\begin{equation}
\label{CW1loop}
%\begin{aligned}
\cW_{\mathrm{(1-loop)}}  \propto 
%(L R)^{-\frac 12} \left(\frac{R}{L}\right)^{D/8}
\cA^{-\frac 12} u^{-\frac D8} 
\left[\eta(\ii u)\right]^{-\frac{D}{2}} \,\ee^{-\sigma \cA}
%\\ & 
=  \cA^{-(\frac 12 + \frac{D}{8})}
\left[\frac{\eta(\ii u)}{\sqrt{R}}\right]^{-\frac{D}{2}} \,
\ee^{-\sigma \cA}~.
%\end{aligned}
\end{equation}
By comparison%
\footnote{Note that in this section we are considering the first order
formulation of the NG action only, without boundary terms, and thus we do not
obtain the perimeter term.}
with \eq{wlngexp}, we deduce that the observable $\cW(L,R)$ of
the first-order formulation
%, that we introduced in \eq{ampli1}, 
is in fact proportional to the Wilson loop $W(L,R)$ through a modular-invariant
prefactor factor of $(LR)^{-\omega} = (LR)^{-(\frac 12 + \frac{D}{8})}$. 
>From the operatorial treatment we extract thus the following prediction for the
Wilson loop up to three loops:
\begin{equation}
\label{wlargefin}
%\begin{aligned}
W(L,R) 
%& 
\propto  \ee^{-\sigma\cA}\,
\left[\frac{\eta(\ii u)}{\sqrt{R}}\right]^{-\frac{D}{2}} \,
%\\ & \times 
\Biggl\{ 1 + \frac{\hat \cL_2(u)}{\sigma\cA}
 + \frac{\hat \cL_3(u)}{(\sigma\cA)^2} + \ldots
\Biggr\}~.
%\end{aligned}
\end{equation}
The two-loop coefficient $\hat\cL_2(u)$ coincides with the (corrected)
expression \eq{2lng} for the two-loop term $\cL_2$, up to the constant
term%
\footnote{In the comparison with the data, the effect of this constant is 
negligible in all cases except the one reported in fig. \ref{fig:results}(d), 
where it enhances the agreement. It would be interesting to understand better
this difference with respect to the Dietz and Filk treatment.}.
This very remarkable and non-trivial  agreement makes us confident in our
result, and in particular in the third loop corrections contained in
eq.s (\ref{wlargefin},\ref{3loopex}). Notice that the overall normalization of
\eq{wlargefin} is totally irrelevant for the observable $\Rw(L,R,n)$ which we
have simulated numerically, since it is given by the ratio of two Wilson loops.

We can also utilize the exact expression
\eq{wlex2} for $\cW(L,R)$, together with the rescaling of the one-loop term
just discussed, to argue that the  Wilson loop can be
written as
\begin{equation}
\label{exW} 
W(L,R) \propto \cA^{\omega} \sum_{k=0}^\infty c_k
\left(\frac{\cE_k}{u}\right)^{\frac D8}\, K_{\frac D8}(\sigma A\cE_k)~.
\end{equation}
This expression can be used to estimate (by truncating the series) the all-loop
prediction of the NG effective string for the Wilson loop and for the
observable $\Rw(L,R,n)$.

\section{Beyond two loops: comparison with the data}
\label{sec:comparison}
\begin{figure}
\centering
\begin{tabular}{cc}
\begin{minipage}{200pt}
\includegraphics[width=1.\textwidth]{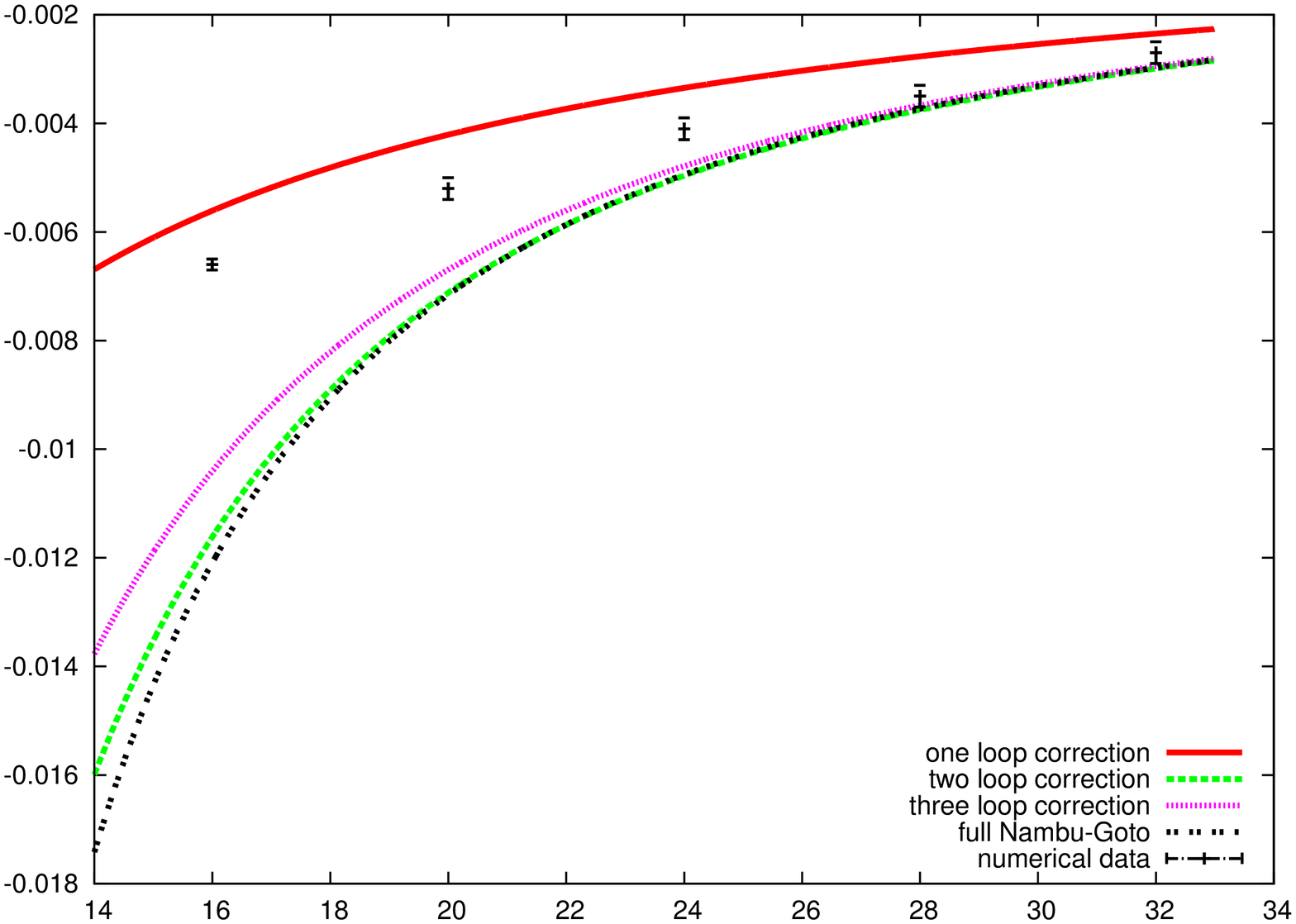}
\caption*{$  \Rw^{'}(L,\frac{3}{4}L,1) $}
\end{minipage} &
\begin{minipage}{200pt}
\includegraphics[width=1.\textwidth]{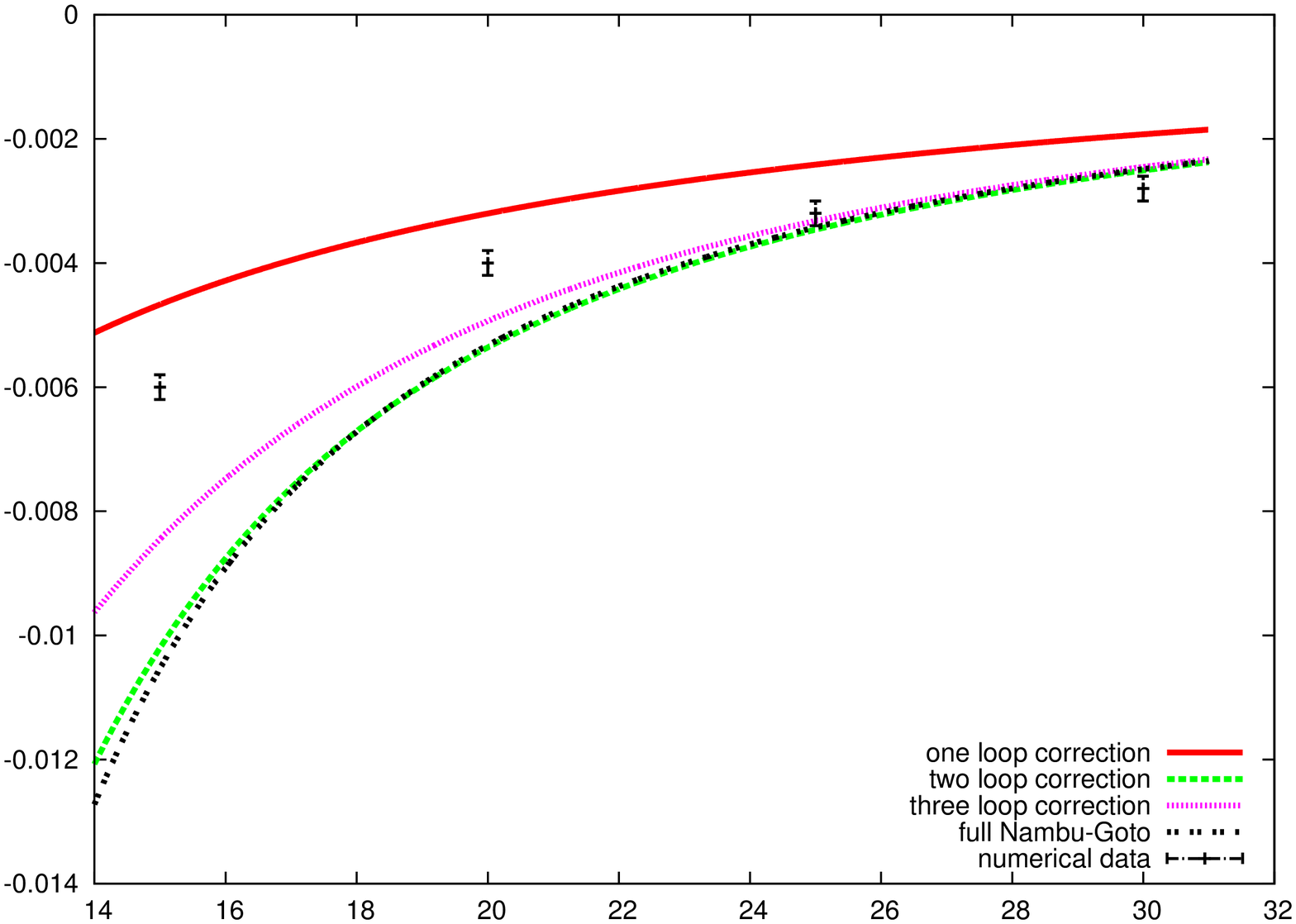}
\caption*{$ \Rw^{'}(L,\frac{4}{5}L,1) $}
\end{minipage} \\
\begin{minipage}{200pt}
\includegraphics[width=1.\textwidth]{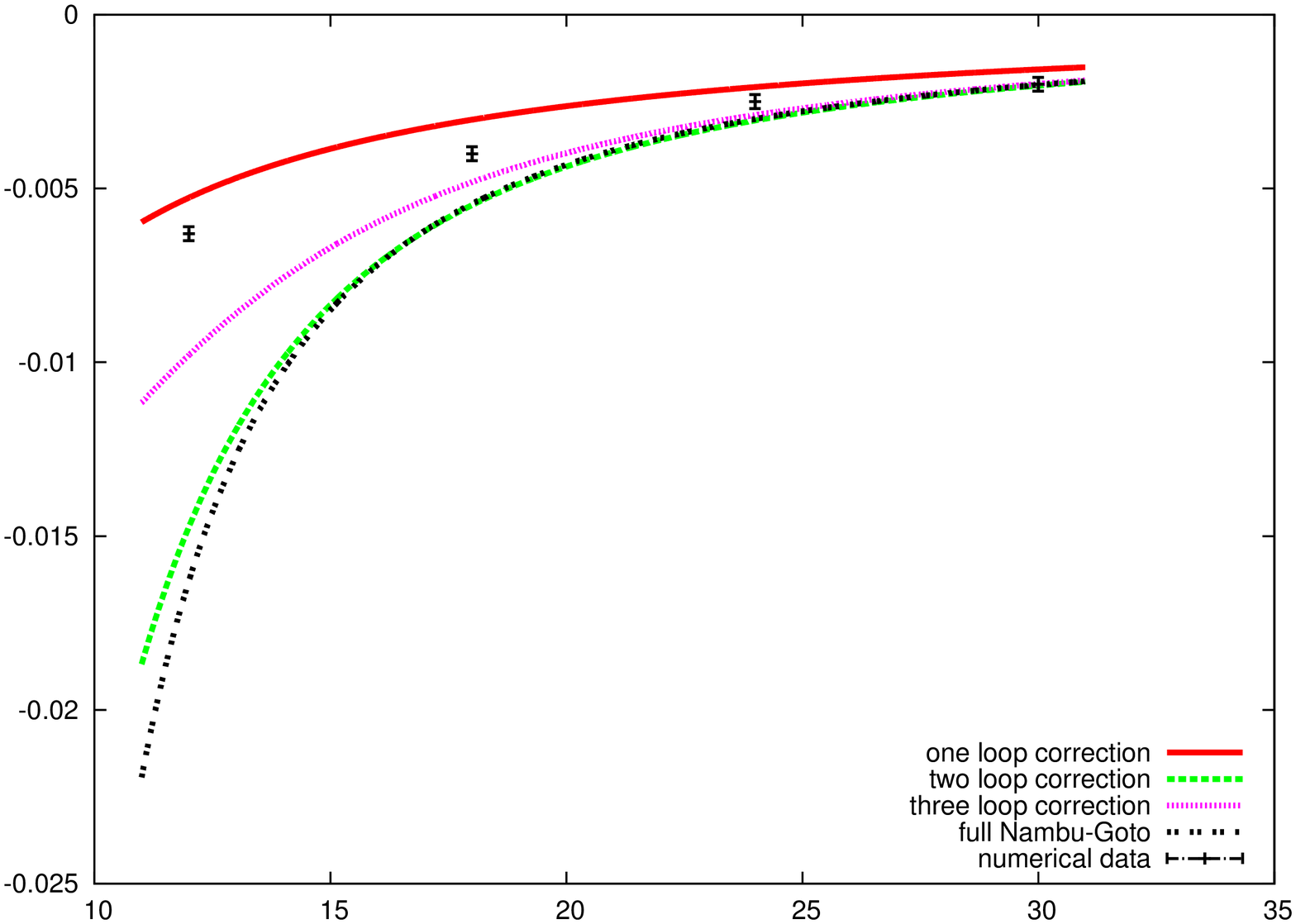}
\caption*{$  \Rw^{'}(L,\frac{5}{6}L,1) $}
\end{minipage} &
\begin{minipage}{200pt}
\includegraphics[width=1.\textwidth]{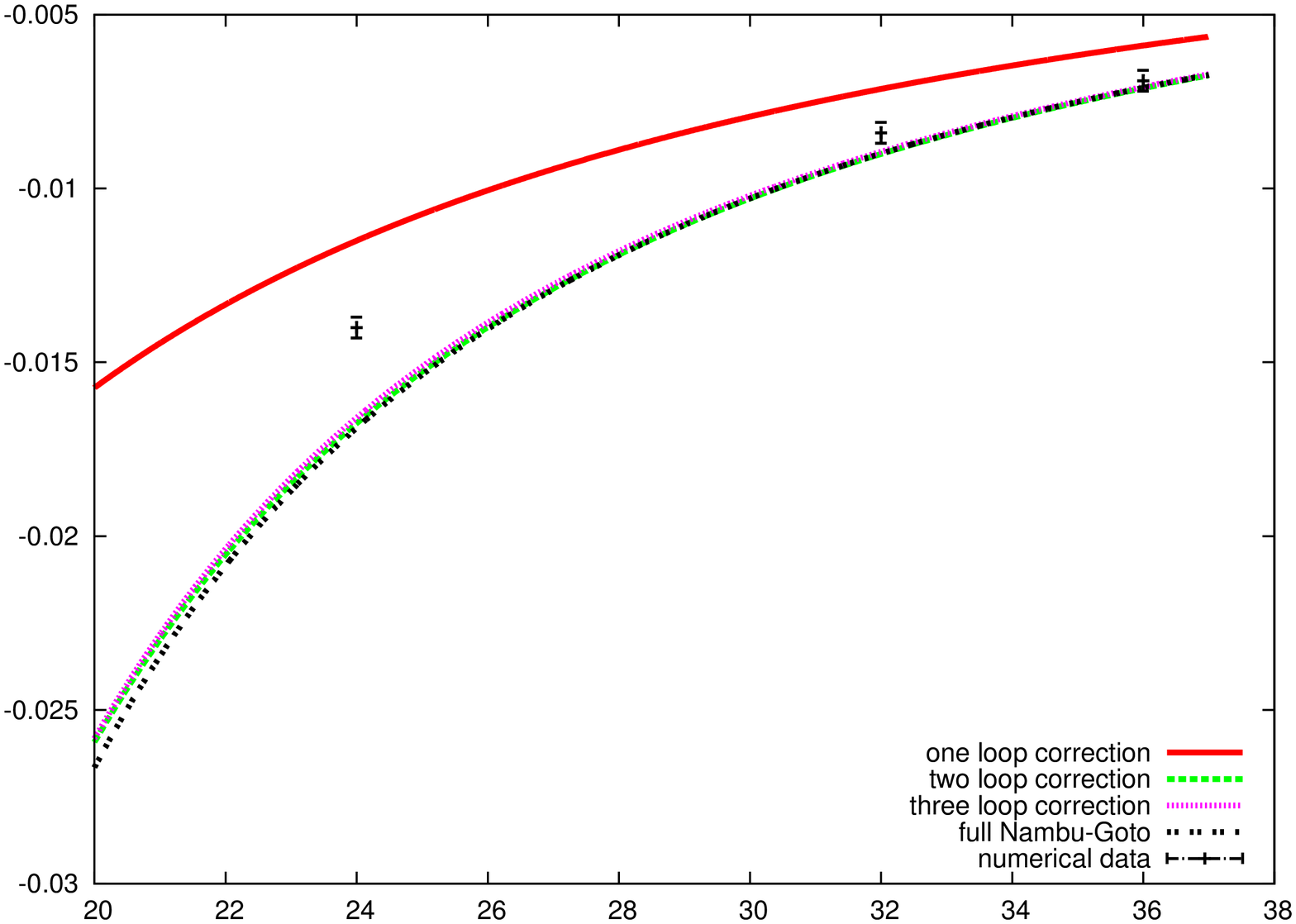}
\caption*{$  \Rw^{'}(L,\frac{3}{4}L,3) $}
\end{minipage}
\end{tabular}
\caption{Plot of $ \Rw^{'}(L,\frac{L}{u},n) $ for various values of $u$ and $n$
against the quantum string corrections up to three loop and the whole
Nambu Goto prediction.}
\label{fig:results3l}
\end{figure}

\begin{figure}
\centering
\begin{tabular}{cc}
\begin{minipage}[t]{200pt}
\includegraphics[width=1.\textwidth]{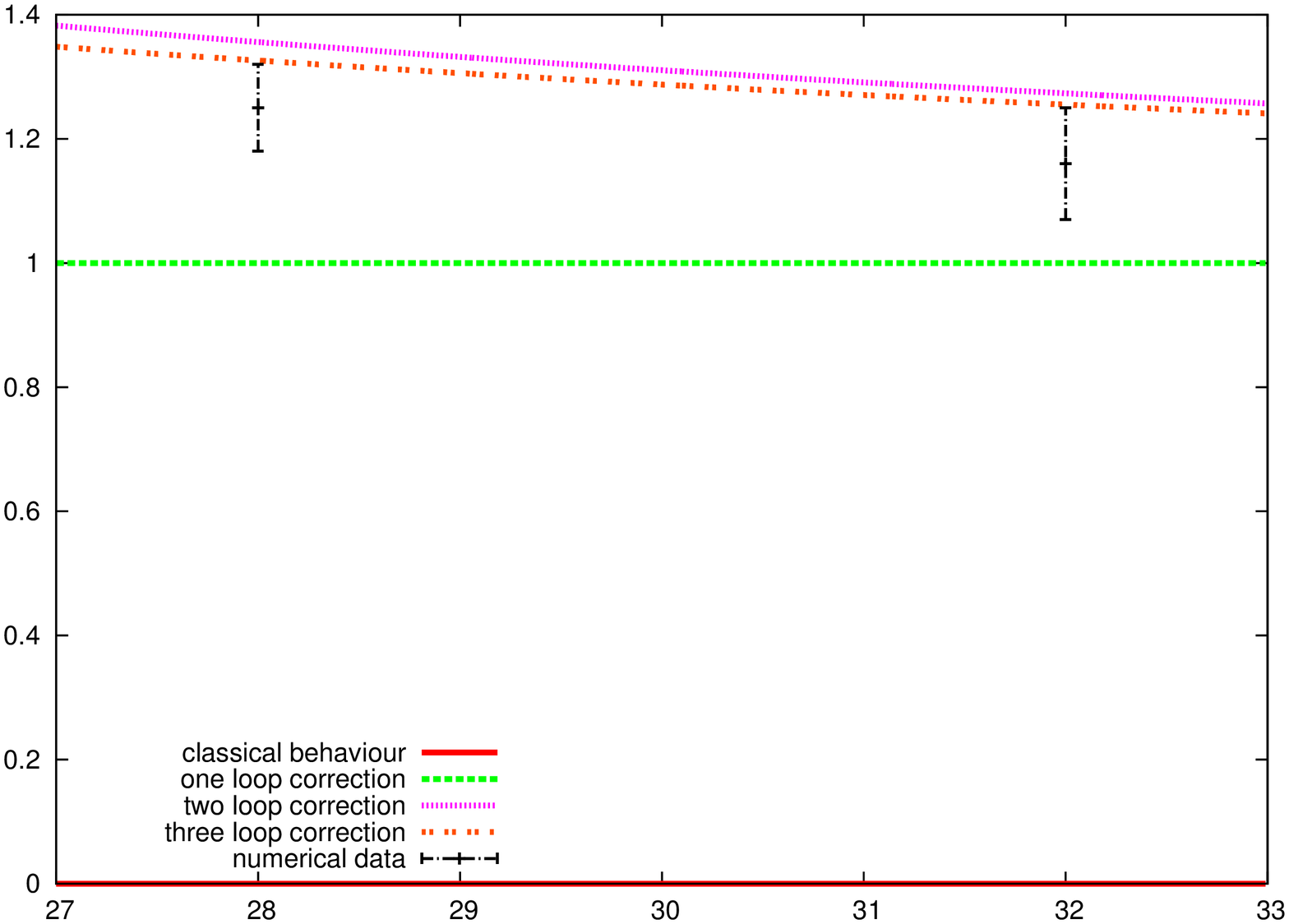}
\caption*{$  \Rw^{'}(L,\frac{3}{4}L,1) $}
\end{minipage} &
\begin{minipage}[t]{200pt}
\includegraphics[width=1.\textwidth]{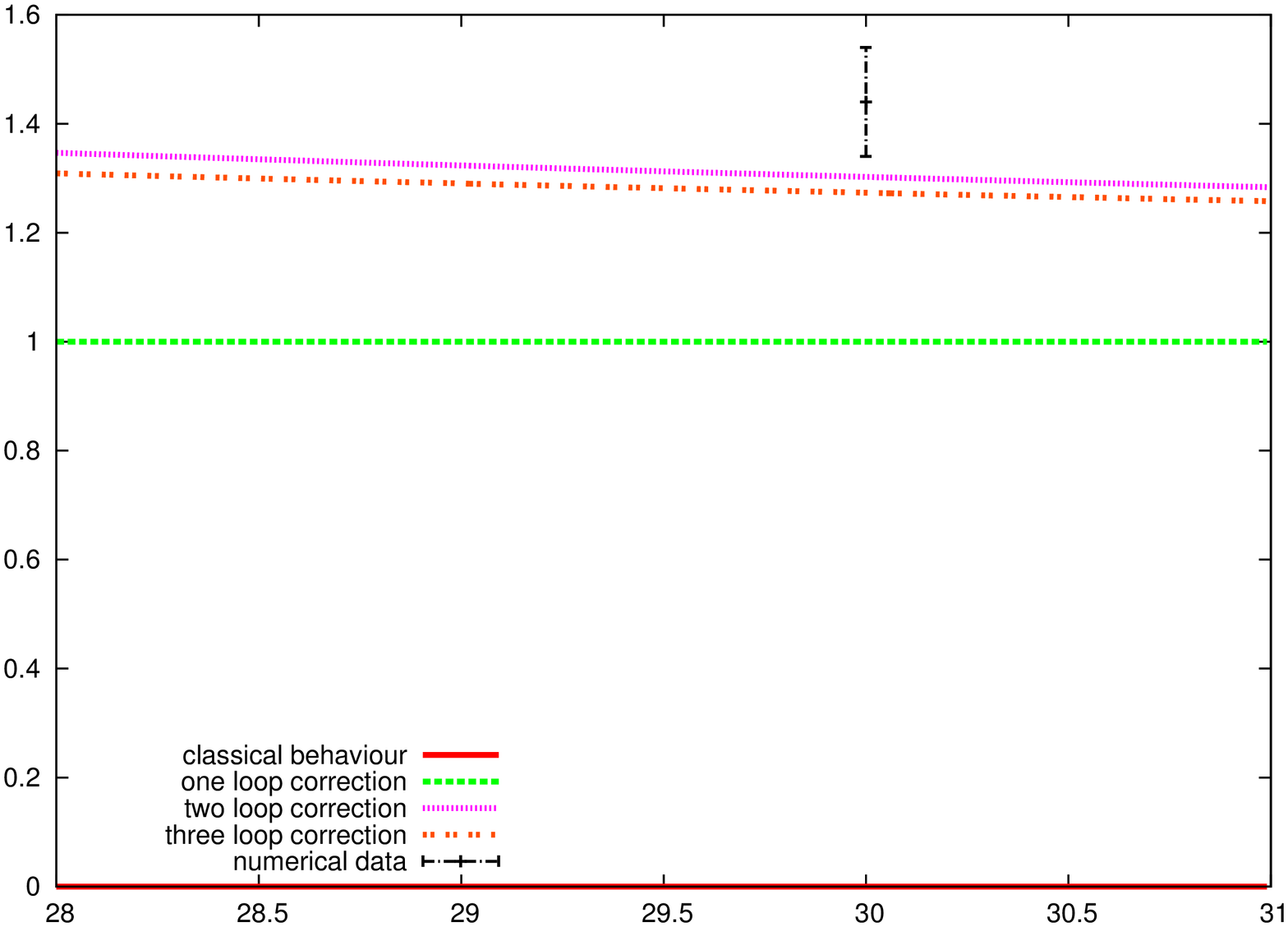}
\caption*{$ \Rw^{'}(L,\frac{4}{5}L,1) $}
\end{minipage} \\
\begin{minipage}[t]{200pt}
\includegraphics[width=1.\textwidth]{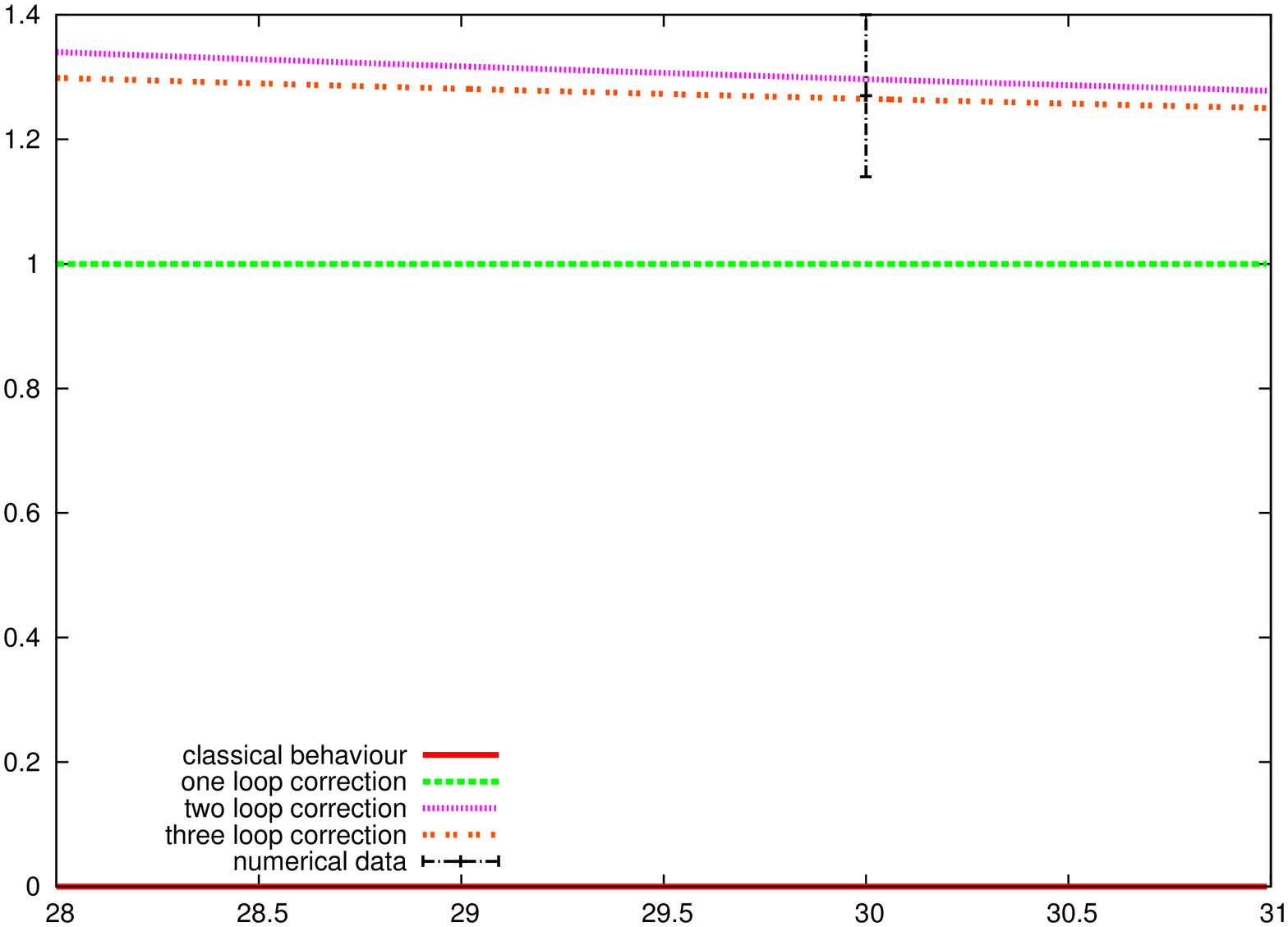}
\caption*{$  \Rw^{'}(L,\frac{5}{6}L,1) $}
\end{minipage} &
\begin{minipage}[t]{200pt}
\includegraphics[width=1.\textwidth]{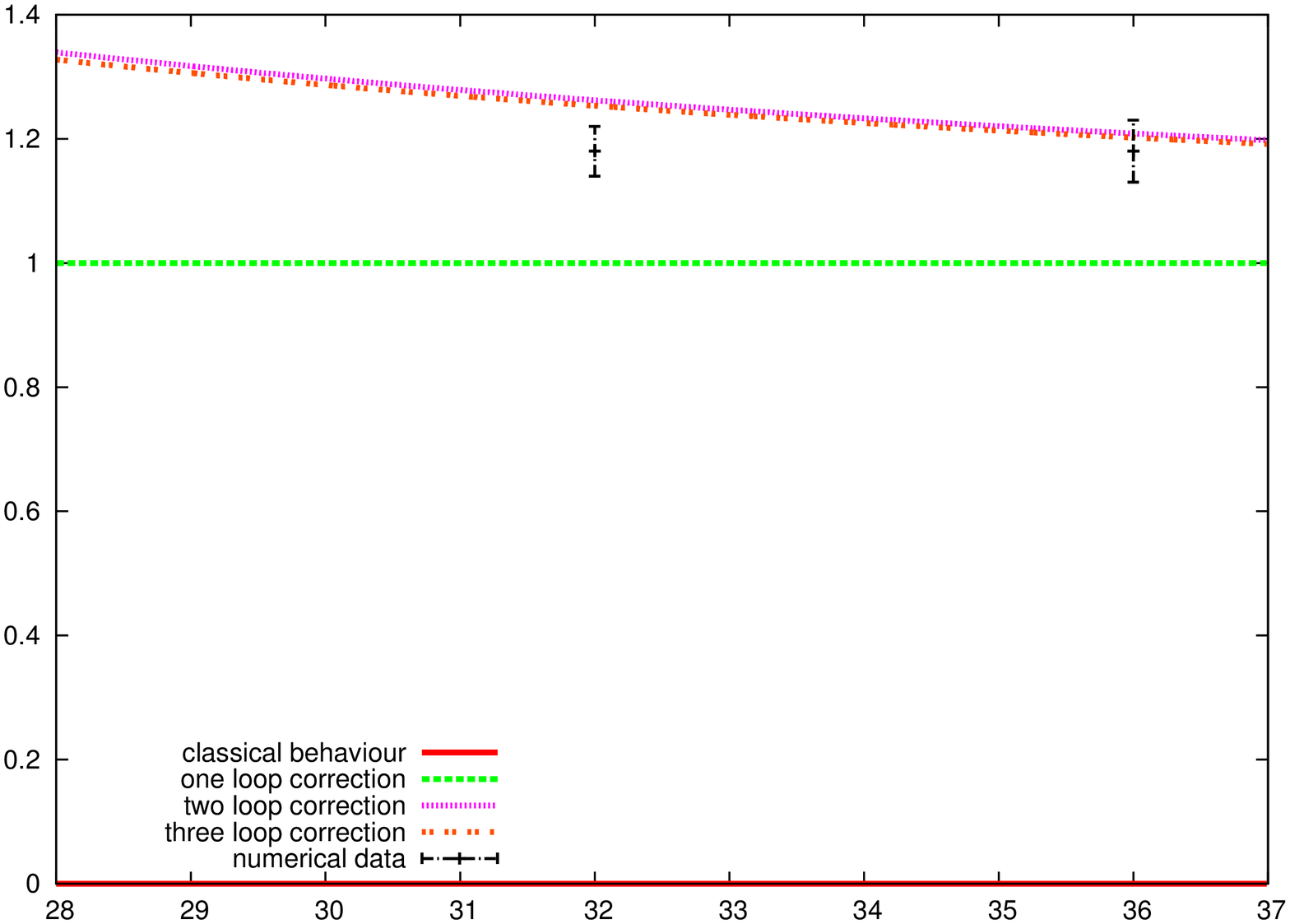}
\caption*{$  \Rw^{'}(L,\frac{3}{4}L,3) $}
\end{minipage}
\end{tabular}
\caption{Same as fig.(\ref{fig:results3l}, but keeping only the data for $R>
20$ which are not affected by boundary conditions (see the main text). 
To make easier the comparison between the predictions at different orders, we 
report here the ratio of $\Rw^{'}$ to the one-loop prediction; the error bars are
rescaled accordingly.}
\label{fig:zoom}
\end{figure}

Using eq.s (\ref{wlargefin},\ref{3loopex}) we may easily extract the three loop
correction for the observable $ \Rw^{'} $ in which we are interested. Similarly,
using eq.(\ref{exW}) we may extract the correction which one would obtain
assuming the validity of the Nambu-Goto action to all orders. These predictions
are reported, together with the numerical data in fig. (\ref{fig:results3l}). To
allow a simpler comparison we reported in fig. (\ref{fig:zoom}) an enlarged
version of the plots, restricted only to the data for $R > 20$ which, as
discussed in section \ref{subsec:comp2l}, are not affected by boundary corrections.

Remarkably enough, the three loop correction turns out to have the opposite sign
with respect to the two loop one and thus it goes exactly in the same direction
as the numerical data. The statistical errors are too large to allow any
reliable test but, as it can be seen in fig. (\ref{fig:results3l}) the gap
between two and three loop corrections increases as $R$ decreases and in
principle, once the boundary corrections will be under control, this particular
combination of Wilson loops could become a perfect setting for high precision
tests of the universality conjectures. In this framework it is also interesting
to notice that this change of sign is somehow peculiar of the three loop
correction and is not present in the whole Nambu-Goto correction (which is also
reported in fig. (\ref{fig:results3l}) for comparison). While a good agreement
of the data with the three loop correction is certainly expected due to
universality, the whole Nambu-Goto action should instead be excluded: it cannot 
be the correct effective action, by the arguments reviewed in the Introduction.
 The data seem indeed to support this statement, even if, as
for the three loop case, the statistical errors are too large to allow any
reliable statement and a definitive answer will require a precise control of the
boundary correction in order to study the corrections in the range $R< 20$.

\section{Conclusions}
\label{sec:conclusions}
In this paper we have considered the disk partition function for the Nambu-Goto
effective string theory corresponding to a rectangular Wilson loop. Our aim was
to check the prediction of the NG model including its quantum corrections up to
three loops against a numerical simulation tailored to this goal. This
purpose, which goes beyond the tests available in the literature, owes its
interest to the theorem which states that the corrections up to three loops are
universal for all effective string models; a check of the NG model to this order
would in fact represent a stringent test of the very idea of an effective string
description.

Setting up such a test required progresses both on the theoretical side and on
the side of simulations.    

On the theoretical side, we have extended to the Wilson loop geometry an
operatorial approach based on the first order re-formulation of the NG model
that had been already used for the cylinder and torus geometry. The operatorial
evaluation makes use of open string analogues of boundary states and leads to an
exact expression. This result resums the loop expansion that arises in the
physical gauge approach, and can be expanded to the desired loop order. In
particular, we have determined the three-loop correction, which had not been
obtained in the physical gauge approach.

On the numerical side, we set up the Montecarlo simulation of an observable
corresponding to the ratio of two Wilson loops. This observable is devised so as
to minimize the effects of boundary terms in the effective action, for which
theoretical predictions are not available. Our simulation has been carried out
in the 3d Ising gauge model, and has reached a level of precision such that it
confirms nicely the validity of the effective string approach up to two loops;
this already represents a big improvement with respect to the avaliable
literature. The statistical errors are yet too large to test reliably the three
loop correction, even though it seems to go in the right direction.

%\vfill \eject
\vskip 1cm

\noindent {\large {\bf Acknowledgments}}
\vskip 0.2cm
We thank F. Gliozzi and I. Pesando for useful discussions. We expecially thank
L. Ferro and V. Verduci, who participated to the early stages of this work. M.B.
thanks the participants to the first week of the "String Phenomenology 2011"
program at NORDITA, and in particular C. Angelantonj, M. Berg, P. Di Vecchia, G.
Ferretti, M. Frau, A. Lerda, A. Mariotti, C. Petersson, R. Russo, A.
Schellekens, for a stimulating discussion session on subjects related to this work. 

\vskip 1cm
\appendix

\section{Notation, conventions and useful formul\ae}
\label{app:not}
Here we establish our notations and give some properties of modular functions 
and Bessel functions that we use in the main text, expecially in section 
\ref{sec:beyond}.
\paragraph{Dedekind function  and Eisenstein series}
Dedekind's $\eta$-function is defined, in terms of the quantity $q=\exp\{2\pi \ii \tau\}$, 
by
\begin{equation} 
\label{defeta1}
\eta(\tau)=q^{\frac{1}{24}} \prod_{n=1}^{\infty}
\left( 1-q^{n} \right)~.
\end{equation}
One can expand its inverse in $q$-series:
\begin{equation}
\label{defeta2}
[\eta(\tau)]^{-1}=\sum_{k=0}^{\infty} p_{k}\, q^{k-\frac{1}{24}}~,
\end{equation}
where $p_{k}$ denotes the number of partitions of $k$. Other powers of 
$\eta$ admit a similar Fourier expansion, and we make use of this fact in \eq{etaexp}. 

Under the generators of modular transformations
% $T$ and $S$ 
the Dedekind eta function transforms in the following way:
\begin{equation}
\label{mteta}
\begin{aligned}
\eta(\tau+1)&= \ee^{\frac{\ii\pi}{12}} \eta(\tau)~, \\
\eta(-1/\tau)&= (\ee^{-\ii\frac{\pi}{2}}\tau)^{\frac{1}{2}} \eta(\tau)~.
\end{aligned}
\end{equation}

Eisenstein series can be defined through their Fourier expansion, which takes the form
\begin{equation}
\label{feeis}
E_{2k}(\tau) = 1 + \frac{2}{\zeta(1-2k)} \sum_{n=1}^\infty
\sigma_{2k-1}(n)\, q^{n}~,
\end{equation} 
where $\sigma_{2k-1}(n)$ is the sum of the $(2k-1)$-th power of the divisors 
of $n$:
\begin{equation}
\label{sigmadiv}
\sigma_\alpha(n) = \sum_{d|n} d^\alpha~.
\end{equation}
In particular, we have thus
\begin{equation}
\label{E23}
\begin{aligned}
E_2(\tau) & = 1 - 24 \sum_{n=1}^\infty \sigma_{1}(n)\, q^{n}~, \\
E_4(\tau) & = 1 + 240 \sum_{n=1}^\infty \sigma_{3}(n)\, q^{n}~,\\
E_6(\tau) & = 1 - 504 \sum_{n=1}^\infty \sigma_{5}(n)\, q^{n}~.
\end{aligned}
\end{equation}

For $k>1$, the Eisenstein series are modular forms of weights 
$2k$:
\begin{equation}
\label{modeis}
%\begin{aligned}
E_{2k}\left(\frac{A\tau + B}{C\tau + D} \right) =
(C\tau + D)^{2k} E_{2k}(\tau)~,
%\\
%E_6\left(\frac{A\tau + B}{C\tau + D} \right) & =
%(C\tau + D)^6 E_6(\tau)~.
%\end{aligned}
\end{equation}
so that in particular
\begin{equation}
 \label{e4e6mod}
E_4(-1/\tau) = \tau^4 E_4(\tau)~,~~~
E_6(-1/\tau) = \tau^6 E_4(\tau)~.
\end{equation}
The series $E_2(\tau)$, instead, is \emph{almost} a modular form of degree 2,
since its transformation under the $S$-generator has a non-homogeneus term:
\begin{equation}
\label{StrE2}
E_2(-\frac{1}{\tau}) = \tau^2 E_2(\tau) - \frac{6}{\pi}\ii \tau~.
\end{equation}
Taking into account these transformation properties, it is not difficult to
check that the various expressions given in section \ref{subsec:wlp}, and in
particular \eq{wlarge3}, are invriant under the exchange $L\leftrightarrow R$,
i.e., $u\leftrightarrow 1/u$.

The $E_2$ series is related to Dedekind $\eta$-function by 
\begin{equation}
\label{E2toeta}\frac{1}{2\pi\ii} \partial_\tau \log\eta(\tau) =
q\partial_q \log\eta(\tau) = \frac{1}{24} E_2(\tau)~. 
\end{equation}
Further derivatives connect $E_2$ to $E_4$ and $E_6$:
\begin{equation}
\label{Eisring}
\begin{aligned}
q\partial_q E_2 & = -\frac{1}{12}(E_4 - E_2^2)~,\\
q\partial_q E_4 & = -\frac{1}{3}(E_6 - E_2 E_4)~,\\
q\partial_q E_6 & = -\frac 12(E_4^2 - E_2 E_6)~.
\end{aligned}
\end{equation}

Applying these formul\ae, we can evaluate the multiple logarithmic derivatives
with respect to $Q=\exp(2\pi\ii u)$ that appear in \eq{wlarge2}:
\begin{equation}
\label{multdereta}
\begin{aligned}
\left(Q\frac{d\phantom{Q}}{dQ}\right) \left[\eta(\ii u)\right]^{-\alpha} & =  
-\frac{\alpha}{24} \left[\eta(\ii u)\right]^{-\alpha} E_2(\ii u)~,\\
\left(Q\frac{d\phantom{Q}}{dQ}\right)^2 \left[\eta(\ii u)\right]^{-\alpha} & = 
\frac{\alpha}{(24)^2} \left[\eta(\ii u)\right]^{-\alpha} \Bigl(2 E_4(\ii u) +
(\alpha -2) E_2^2(\ii u)\Bigr)~,\\
\left(Q\frac{d\phantom{Q}}{dQ}\right)^3 \left[\eta(\ii u)\right]^{-\alpha} & = 
-\frac{\alpha}{(24)^3} \left[\eta(\ii u)\right]^{-\alpha} \Bigl(16 E_6(\ii u) +
6(\alpha - 4)E_4(\ii u) E_2(\ii u)\\
& + (\alpha^2 - 6 \alpha + 8) E_2^3(\ii u)\Bigr)~,\\
\left(Q\frac{d\phantom{Q}}{dQ}\right)^4 \left[\eta(\ii u)\right]^{-\alpha} & = 
\frac{\alpha}{(24)^4} \left[\eta(\ii u)\right]^{-\alpha} \Bigl(64(\alpha - 6)
E_6(\ii u) E_2(\ii u)
+ 12 (\alpha + 12) E_4^2(\ii u)\\
& + 12 (\alpha^2 - 10 \alpha + 24) E_4(\ii u) E_2^2(\ii u) + (\alpha^3 - 12
\alpha^2 + 44\alpha -48) E_2^4(\ii u)\Bigr)~.
\end{aligned}
\end{equation}

\paragraph{Asymptotic expansion of Bessel Functions}
The asympotic expansion of the modified Bessel functions of the second kind
$K_\alpha(z)$ for large values of their argument $z$ takes the form
\begin{equation}
 \label{asbessel}
K_\alpha(z) \sim \sqrt{\frac{\pi}{2 z}} \ee^{-z}\sum_m
\frac{1}{m!}\frac{\textstyle\Gamma(\alpha + m + \frac
12)}{\textstyle\Gamma(\alpha - m + \frac 12)} \frac{1}{(2 z)^m}~.
\end{equation}

%\providecommand{\href}[2]{#2}\begingroup\raggedright
%\bibliography{references_effstring}
%\bibliographystyle{abe}

\providecommand{\href}[2]{#2}\begingroup\raggedright\endgroup

\end{document}